\documentstyle[preprint,aps,epsfig]{revtex}

\tightenlines

\begin{document}
                  
\title{Electromagnetic Structure of the Nucleon \\
in the Perturbative Chiral Quark Model} 

\author{V. \ E. \ Lyubovitskij \footnotemark[1]\footnotemark[2], 
Th. \ Gutsche \footnotemark[1] and Amand Faessler \footnotemark[1]
\vspace*{0.4\baselineskip}}
\address{
\footnotemark[1] 
Institut f\"ur Theoretische Physik, Universit\"at 
T\"ubingen, Auf der Morgenstelle 14,  \\
D-72076 T\"ubingen, Germany 
\vspace*{0.2\baselineskip}\\
\footnotemark[2]
Department of Physics, Tomsk State University, 634050 Tomsk, Russia 
\vspace*{0.3\baselineskip}\\} 

\maketitle

\vskip.5cm

\begin{abstract}
We apply the perturbative chiral quark model (PCQM) at one loop to 
analyse the electromagnetic structure of nucleons. This model is 
based on an effective Lagrangian, where baryons are described by 
relativistic valence quarks and a perturbative cloud of Goldstone 
bosons. Including the electromagnetic interaction we first develop 
the formalism up to one-loop in the Goldstone boson fluctuation 
relying on renormalization by use of counterterms. 
Local gauge invariance is satisfied both on the Lagrangian level and 
also for the relevant baryon matrix elements in the Breit frame. 
We apply the formalism to obtain analytical expressions for the nucleon 
charge and magnetic form factors, which are expressed in terms of 
fundamental parameters of low-energy pion-nucleon physics (weak pion 
decay constant, axial nucleon coupling, strong pion-nucleon form factor) 
and of only one model parameter (radius of the nucleonic three-quark core). 
A detailed numerical analysis for the nucleon magnetic moments, charge and 
magnetic radii and also for the momentum dependence of form factors 
is presented. 
\end{abstract}

\vspace*{\baselineskip}
\vskip1cm

\noindent {\it PACS:} 
11.10.Ef, 12.39.Fe, 12.39.Ki, 13.40.Em,13.40.Gp, 14.20.Dh  
       
\vskip.5cm

\noindent {\it Keywords:} Chiral symmetry; Relativistic quark model; 
Relativistic effective Lagrangian; Nucleon electromagnetic form factors.

\section{Introduction} 

The study of the electromagnetic form factors of the nucleon is of fundamental 
importance in hadron physics to gain a deeper understanding of the baryon 
structure and the interplay between strong and electromagnetic interactions. 
Completed and ongoing experiments at ELSA, JLab, MAMI, MIT, NIKHEF and other 
laboratories on improved measurements of the elastic proton and neutron form 
factors stimulated a comprehensive theoretical study of these quantities. 
The theoretical description of electromagnetic form factors was performed in 
detail within different approaches of low-energy hadron physics: QCD Sum 
Rules, Chiral Perturbation Theory, relativistic and non-relativistic 
quark models, QCD-motivated approaches based on solutions of the 
Schwinger-Dyson/Bethe-Salpeter and Faddeev equations, soliton-type models, 
etc. For recent experimental and theoretical advances with respect to the 
electromagnetic structure of nucleons see the proceedings of the last 
conferences \cite{MENU99}-\cite{BARYON98}.  

Since the early eighties chiral quark models 
\cite{Theberge-Thomas}-\cite{Gutsche}, describing the
nucleon as a bound system of valence quarks with a surrounding
pion cloud, play an important role in the description of low-energy nucleon
physics. These models include the two main features of low-energy hadron
structure, confinement and chiral symmetry. With respect to the treatment
of the pion cloud these approaches fall essentially into two categories.
 
The first type of chiral quark models assumes that the valence quark
content dominates the nucleon, thereby treating pion contributions
perturbatively \cite{Theberge-Thomas,Oset,Chin,Gutsche}. 
Originally, this idea was formulated in the context of the cloudy bag
model \cite{Theberge-Thomas}. By imposing chiral symmetry the MIT
bag model \cite{MIT} was extended to include the interaction
of the confined quarks with the pion fields on the bag surface.
With the pion cloud treated as a perturbation on the basic features of
the MIT bag, pionic effects generally improve the
description of nucleon observables.
Later, similar perturbative chiral models \cite{Oset,Chin,Gutsche} were 
developed where the rather unphysical sharp bag boundary is replaced by 
a finite surface thickness of the quark core. 
By introducing a static quark potential of general form, these quark models
contain a set of free parameters characterizing the confinement (coupling
strength) and/or the quark masses. The perturbative technique allows a fully
quantized treatment of the pion field up to a given order in accuracy.
Although formulated on the quark level, where confinement is put in
phenomenologically, perturbative chiral quark models are formally close
to chiral perturbation theory on the hadron level.

Alternatively, when the pion cloud is assumed to dominate the nucleon 
structure this effect has to be treated non-perturbatively. 
The non-perturbative approaches are based for example on 
Refs. \cite{Diakonov-Petrov}, where the chiral quark soliton model was 
derived. This model is based on the concept that the QCD instanton vacuum 
is responsible for the spontaneous breaking of chiral symmetry, which in 
turn leads to an effective chiral Lagrangian at low energy as "derived"
from QCD.
The classical pion field (the soliton) is described by a trial profile
function, which is fixed by minimizing the energy of the nucleon.  
Further quantization of slow rotations of this soliton field leads to a
nucleon state, which is built up from rotational excitations of
the classical nucleon.
On the phenomenological level the chiral quark soliton model tends to
be advantageous in the description of the nucleon spin structure, that is
for large momentum transfers, but is comparable to the original perturbative
chiral quark models in the description of low-energy nucleon properties.

As a further development of chiral quark models with a perturbative
treatment of the pion cloud \cite{Theberge-Thomas,Oset,Chin,Gutsche}, 
we extended the relativistic quark model suggested in \cite{Gutsche}
for the study of the low-energy properties of the nucleon \cite{PCQM}.
Compared to the previous similar models of 
Refs. \cite{Theberge-Thomas,Oset,Chin} our current approach contains 
several new features: i) generalization of the phenomenological 
confining potential; ii) SU(3) extension of chiral symmetry to include
the kaon and eta-meson cloud contributions; iii) consistent formulation of
perturbation theory both on the quark and baryon level by use of
renormalization techniques and by allowing to account for excited quark
states in the meson loop diagrams; iv) fulfillment of the
constraints imposed by chiral symmetry (low-energy theorems),
including the current quark mass expansion of the matrix elements 
(for details see Ref. \cite{PCQM}); v) possible consistency with chiral 
perturbation theory as for example demonstrated \cite{PCQM} for the chiral 
expansion of the nucleon mass. In the following we refer to our model as 
the {\it perturbative chiral quark model} (PCQM).

The PCQM is based on an effective chiral Lagrangian 
describing quarks as relativistic fermions moving in a self-consistent 
field (static potential). The latter is described by a scalar potential 
$S$ providing confinement of quarks and the time component of a vector 
potential $\gamma^0 V$ responsible for short-range fluctuations of the 
gluon field configurations \cite{Luscher} (see also recent lattice 
calculations \cite{Takahashi}). The model potential defines unperturbed 
wave functions of quarks which are subsequently used in the calculation 
of baryon properties. Baryons in the PCQM are described as bound states 
of valence quarks surrounded by a cloud of Goldstone bosons 
($\pi, K, \eta$) as required by chiral symmetry. Interaction of quarks 
with Goldstone bosons is introduced on the basis of the nonlinear 
$\sigma$-model \cite{Gell-Mann_Levy}. When considering mesons fields as 
small fluctuations we restrict ourselves to the linear form of the 
meson-quark interaction. With the derived interaction Lagrangian we do 
our perturbation theory in the expansion parameter $1/F$ (where $F$ is 
the pion leptonic decay constant in the chiral limit). We also treat the 
mass term of the current quarks as a small perturbation. Dressing the 
baryon three-quark core by a cloud of Goldstone mesons corresponds to the 
inclusion of the sea-quark contribution. All calculations are performed 
at one loop or at order of accuracy $o(1/F^2, \hat m, m_s)$. 
The chiral limit with $\hat m, m_s \to 0$ is well defined. 

In Ref. \cite{PCQM} our approach was successfully applied to the study of 
sigma-term physics. Our result for the $\pi N$ sigma term 
$\sigma_{\pi N} \approx$ 45 MeV is in good agreement with the value deduced 
by Gasser, Leutwyler and Sainio \cite{GLS} using dispersion-relation 
techniques and exploiting the chiral symmetry constraints. 
A complete analysis of meson-nucleon and meson-delta-isobar sigma-terms, 
is given including the strangeness content of the nucleon. 
We have a physically reasonable prediction for the quantity 
$y_N = 0.076 \pm 0.012$ \cite{PCQM} characterizing the strange quark 
contribution to the nucleon mass. Our predictions 
for the strangeness content of the nucleon and for $KN$ sigma-terms are 
important for the ongoing DA$\Phi$NE experiments at Frascati \cite{Gensini}. 

In this paper, we apply the PCQM  to analyse the electromagnetic structure 
of nucleons. The main intention of the present work is to develop the full 
formalism for the effective Lagrangian of the PCQM, where Goldstone boson 
fluctuations are treated as higher order effects to the pure valence quark 
core results. We give detailed consideration of gauge invariance in 
noncovariant chiral quark model treating pion cloud perturbatively.  

In the present article we proceed as follows. In the following section, we 
first describe the basic notions of our approach: the underlying effective 
Lagrangian, choice of parameters and fulfilment of low-energy theorems. 
We extend our model Lagangian by including the electromagnetic interaction. 
A consistent formalism of the perturbation series up to one meson loop and up 
to terms linear in the current quark masses is presented. We utilize 
a renormalization technique, which, by introducing counterterms, greatly 
simplifies the evaluation. Thereby, local gauge invariance of the 
electromagnetic interaction is shown to be fulfilled both on the Lagrangian 
level and for baryon matrix elements set up in the Breit frame. In Sect. 3 
we concentrate on the specific application of the PCQM to the electromagnetic 
properties of the nucleon. We derive analytical expressions for 
the nucleon charge and magnetic form factors expressed in terms of 
fundamental parameters of low-energy pion-nucleon physics (weak pion decay 
constant, axial nucleon coupling, strong pion-nucleon form factor) and of 
only one model parameter (radius of the nucleonic three-quark core). 
Numerical results in comparison with data are presented to test the basic and 
standard phenomenological implications of the model. Finally, Sect. 4 contains 
a summary of our major conclusions. 

\section{Perturbative chiral quark model} 

\subsection{Effective Lagrangian and zeroth order properties}

The perturbative chiral quark model (PCQM) \cite{PCQM} is based on an 
effective chiral Lagrangian describing the valence quarks of baryons as 
relativistic fermions moving in a self-consistent field (static potential) 
$V_{eff}(r)=S(r)+\gamma^0 V(r)$ with $r=|\vec{x}|$ 
\cite{Gutsche,Oset,Fetter_Walecka}, which are supplemented by a cloud of 
Goldstone bosons $(\pi, K, \eta)$. Treating Goldstone fields as small 
fluctuations around the three-quark (3q) core we derive a linearized 
effective Lagrangian ${\cal L}_{eff}$. The Lagrangian 
${\cal L}_{eff}={\cal L}_{inv}^{lin}+{\cal L}_{\chi SB}$, 
derived in Ref. \cite{PCQM}, includes the linear chiral-invariant term 
\begin{eqnarray}\label{Lagrangian_lin_inv}
{\cal L}_{inv}^{lin}(x)=\bar\psi(x)[i\not\!\partial-S(r)-\gamma^0V(r)]\psi(x) 
\,+\,\frac{1}{2} (\partial_\mu\hat{\Phi}(x))^2
- \bar\psi(x)S(r)i\gamma^5 \frac{\hat\Phi(x)}{F} \psi(x)   
\end{eqnarray}
and a term ${\cal L}_{\chi SB}$ which explicitly breaks chiral symmetry 
\begin{eqnarray}\label{Lchi}
{\cal L}_{\chi SB}(x)=-\bar\psi(x){\cal M}\psi(x) 
-\frac{B}{8} {\rm Tr}\{ \hat\Phi(x) , \, \{ \hat\Phi(x) , \, {\cal M} \}\}  
\end{eqnarray}
containing the mass terms for quarks and mesons \cite{PCQM}. 
The octet matrix  $\hat\Phi$ of pseudoscalar mesons is defined as:  
\begin{equation}
\frac{\hat\Phi}{\sqrt{2}}=
\sum_{i=1}^{8}\frac{\Phi_i\lambda_i}{\sqrt{2}}=
\left(
\begin{array}{ccc}
\pi^0/\sqrt{2} + \eta/\sqrt{6}\,\, & \,\, \pi^+ \,\, & \, K^+ \\
\pi^- \,\, & \,\, -\pi^0/\sqrt{2}+\eta/\sqrt{6}\,\, & \, K^0\\
K^-\,\, & \,\, \bar K^0 \,\, & \, -2\eta/\sqrt{6}\\
\end{array}
\right);   
\end{equation} 
$F = 88$ MeV \cite{PCQM,Gasser1} is the pion decay constant in the chiral 
limit; ${\cal M} = {\rm diag}\{\hat m, \hat m, m_s\}$ is the mass matrix of 
current quarks\footnote{Here we restrict to the isospin symmetry limit with 
$m_u = m_d = \hat m$.}; $B=-<0|\bar u u|0>/F^2$ is the low-energy constant 
which measures the vacuum expectation value of the scalar quark densities in 
the chiral limit~\cite{Gasser_Leutwyler}. We rely on the standard picture of 
chiral symmetry breaking~\cite{Gasser_Leutwyler} and for the masses of 
pseudoscalar mesons we use the leading term in their chiral expansion 
(i.e. linear in the current quark mass):  
\begin{eqnarray}\label{M_Masses}
M_{\pi}^2=2 \hat m B, \hspace*{.5cm} M_{K}^2=(\hat m + m_s) B, \hspace*{.5cm} 
M_{\eta}^2= \frac{2}{3} (\hat m + 2m_s) B. 
\end{eqnarray}
Meson masses obviously satisfy the Gell-Mann-Oakes-Renner (\ref{M_Masses})  
and the Gell-Mann-Okubo relation $3 M_{\eta}^2 + M_{\pi}^2 = 4 M_{K}^2$. 
In the evaluation we use the following set of QCD 
parameters~\cite{Gasser_Leutwyler_PR}:  
\begin{eqnarray}
\hat m= 7 \,\, \mbox{MeV}, \,\,\,\,\, \frac{m_s}{\hat m}=25 \,\,\,\,\, 
\mbox{and} \,\,\,\,\, B=\frac{M_{\pi^+}^2}{2\hat m}=1.4 \,\, \mbox{GeV} .
\end{eqnarray}
Furthermore, the linearized effective Lagrangian fulfils the PCAC 
requirement \cite{PCQM}, consistent with the Goldberger-Treiman relation 
(as shown later on).  

To derive the properties of baryons, which are modelled as bound states of 
valence quarks surrounded by a meson cloud, we formulate perturbation theory. 
At zeroth order, the unperturbed Lagrangian simply describes the nucleon by 
three relativistic valence quarks which are confined by the effective one-body 
potential in the Dirac equation. The mass $m_N^{core}$ of the three-quark core 
of the nucleon is then related to the single quark ground state energy 
${\cal E}_0$ by $m_N^{core}=3\cdot {\cal E}_0$ (Here we do not discuss the 
removal of the spurious contribution to the baryon mass arising from the 
centre-of-mass motion of the bound state. It will be subject of a future 
publication.). For the unperturbed three-quark state we introduce the 
notation $|\phi_0>$ with the normalization $<\phi_0|\phi_0>=1$. The single 
quark ground state energy ${\cal E}_0$ and wave function (w.f.) $u_0(\vec{x})$ 
are obtained from the Dirac equation 
\begin{eqnarray}\label{Dirac_Eq}
[-i\vec{\alpha}\vec{\nabla}+\beta S(r)+V(r)-{\cal E}_0]u_0(\vec{x})=0 .
\end{eqnarray}    
The quark w.f. $u_0(\vec{x})$ belongs to the basis of potential 
eigenstates (including excited quark and antiquark solutions) used for 
expanding the quark field operator $\psi(x)$. Here we restrict the  
expansion to the ground state contribution with 
\begin{eqnarray}\label{psi_bare}
\psi(x)=b_0 u_0(\vec{x}) \exp(-i {\cal E}_0 t), 
\end{eqnarray}
where $b_0$ is the corresponding single quark annihilation operator. 

At the unperturbed level the current quark mass is not included in 
Eq. (\ref{Dirac_Eq}) to simplify our calculational technique. Instead we 
consider the quark mass term as a small 
perturbation \cite{Gasser_Leutwyler_PR}, as will be discussed later on. 
Inclusion of a finite current quark mass leads to a displacement of the 
single quark energy which, for example, is relevant for the calculation of 
the sigma-term (see discussion in Ref. \cite{PCQM}); a quantity which 
vanishes in the chiral limit. On the other hand, the effect of a finite 
current quark mass on observables which survive in the chiral limit, like 
magnetic moments, charge radii, etc., is quite negligible. 

For a given form of the potentials $S(r)$ and $V(r)$ the Dirac 
equation (\ref{Dirac_Eq}) can be solved numerically. Here, for the sake of 
simplicity, we use a variational {\it Gaussian ansatz} \cite{Duck} for the 
quark wave function given by the analytical form: 
\begin{eqnarray}\label{Gaussian_Ansatz} 
u_0(\vec{x}) \, = \, N \, \exp\biggl[-\frac{\vec{x}^{\, 2}}{2R^2}\biggr] \, 
\left(
\begin{array}{c}
1\\
i \rho \, \frac{\displaystyle{\vec{\sigma}\vec{x}}}{\displaystyle{R}}\\
\end{array} 
\right) \, \chi_s \, \chi_f\, \chi_c \, , 
\end{eqnarray}      
where 
\begin{eqnarray}
N=[\pi^{3/2} R^3 (1+3\rho^2/2)]^{-1/2}
\end{eqnarray}
is a constant fixed by the normalization condition 
\begin{eqnarray}
\int d^3x \, u^\dagger_0(x) \, u_0(x) \equiv 1;
\end{eqnarray} 
$\chi_s$, $\chi_f$, $\chi_c$ are the spin, flavor and color quark wave 
functions, respectively. Our Gaussian ansatz contains two model parameters: 
the dimensional parameter $R$ and the dimensionless parameter $\rho$. 
The parameter $\rho$ can be related to the axial coupling constant $g_A$ 
calculated in zeroth-order (or 3q-core) approximation: 
\begin{eqnarray}\label{g_A}
g_A=\frac{5}{3} \biggl(1 - \frac{2\rho^2} {1+\frac{3}{2} \rho^2} \biggr) = 
\frac{5}{3} \frac{1+2\gamma}{3} ,  
\end{eqnarray}
where $\gamma$ is a relativistic reduction factor  
\begin{eqnarray}
\gamma=\frac{1-\frac{3}{2}\rho^2}{1+\frac{3}{2}\rho^2}=
\frac{9}{10}g_A-\frac{1}{2} .
\end{eqnarray}
Note that since PCAC is fulfilled in our model \cite{PCQM}, on the tree level 
the axial charge $g_A$ is on the tree level related to the pion-nucleon 
coupling constant $G_{\pi NN}$ by the Goldberger-Treiman relation 
\begin{eqnarray}
G_{\pi NN}=\frac{5m_N}{3F} \, \biggl(1 \, - \, \frac{2\rho^2}
{1 \, + \, \frac{3}{2} \, \rho^2}\biggr) 
\equiv \frac{m_N}{F} \, g_A .  
\end{eqnarray}
where $m_N$ is the physical nucleon mass. 

The parameter $R$ can be physically understood as the mean radius of the 
three-quark core and is related to the charge radius $<r^2_E>^P_{LO}$ 
of the proton in the leading-order (or zeroth-order) approximation 
as \cite{PCQM} 
\begin{eqnarray}\label{r2ep_LO}
<r^2_E>^P_{LO} \, = \, \frac{3R^2}{2} \, 
\frac{1 \, + \, \frac{5}{2} \, \rho^2}
{1 \, + \, \frac{3}{2} \, \rho^2} \, = \, 
R^2 \biggl( 2 - \frac{\gamma}{2} \biggr) . 
\end{eqnarray}
In our calculations we use the value $g_A$=1.25 as obtained in 
ChPT \cite{Gasser1}. We therefore have only one free parameter, $R$. 
In the numerical evaluation $R$ is varied in the region from 0.55 fm  
to 0.65 fm corresponding to a change of $<r^2_E>^P_{LO}$ in the 
region from 0.5 fm$^2$ to 0.7 fm$^2$. The use of the Gaussian ansatz 
(\ref{Gaussian_Ansatz}) in its exact form restricts the scalar confinement 
potential $S(r)$ to  
\begin{eqnarray}\label{Pot-Sr}
S(r) = \underbrace{\frac{1 \, - \, 3\rho^2}{2 \, \rho R}}_{= \, M} 
\,\, + \,\,
\underbrace{\frac{\rho}{2R^3}}_{= \, c} \, r^2 = M \,\, + \,\, c \, r^2,   
\end{eqnarray}
expressed in terms of the parameters $R$ and $\rho$. The constant part of the 
scalar potential $M$ can be interpreted as the constituent mass of the quark, 
which is simply a displacement of the current quark mass due to the potential 
$S(r)$. The parameter $c$ is a constant defining the radial (quadratic) 
dependence of the scalar potential. Numerically, for our set of parameters, 
we get $M=230\pm 20$ MeV and $c=0.007\pm 0.002$ GeV$^3$. Error bars for 
theoretical results quoted here and in the following correspond to the 
variation of the parameter $R$. For the vector potential we get the following 
expression 
\begin{eqnarray}
V(r) = {\cal E}_0 \,\,  - \,\, \frac{1 \, + \, 3\rho^2}{2 \, \rho R}
\,\, + \,\, \frac{\rho}{2R^3} \, r^2, 
\end{eqnarray}
where the single quark energy ${\cal E}_0$ is a free parameter in the 
Gaussian ansatz. 

\subsection{Perturbation theory and nucleon mass}

Before setting out to present the renormalization scheme of the PCQM, we 
first define and discuss the quantities, relevant for mass and wave function 
renormalization. Following the Gell-Mann and Low theorem \cite{Gell-Mann_Low} 
we define the mass shift of the nucleonic three-quark ground state 
$\Delta m_N$ due to the interaction with Goldstone mesons as\footnote{We do 
not take into account the electromagnetic contribution to the nucleon mass 
since this effect is quite negligible.} 
\begin{eqnarray}\label{Energy_shift} 
\Delta m_N \, = \, ^N\!\!<\phi_0| \, 
\sum\limits_{n=1}^{\infty} \frac{i^n}{n!} \, 
\int \, i\delta(t_1) \, d^4x_1 \ldots d^4x_n \, 
T[{\cal L}_I(x_1) \ldots {\cal L}_I(x_n)] \, |\phi_0>_{c}^{N} .   
\end{eqnarray} 
In Eq. (\ref{Energy_shift}) the strong interaction Lagrangian ${\cal L}_I$ 
treated as a perturbation is defined as\footnote{In the calculation of matrix 
elements we use the interaction Lagrangian and Wick's $T$-ordering product   
for field operators \cite{Bogoliubov_Shirkov}.}   
\begin{eqnarray}\label{L_I} 
{\cal L}_I(x) = - \bar\psi(x) i\gamma^5 \frac{\hat\Phi(x)}{F} S(r) \psi(x) 
\end{eqnarray} 
and subscript $"c"$ refers to contributions from connected graphs only. We 
evaluate Eq. (\ref{Energy_shift}) at one loop to order $o(1/F^2)$ using Wick's 
theorem and the appropriate propagators. For the quark field we use a Feynman 
propagator for a fermion in a binding potential. By restricting the summation 
over intermediate quark states to the ground state we get   
\begin{eqnarray} 
\hspace*{-.75cm} 
iG_\psi(x,y)=<\phi_0|T\{\psi(x)\bar\psi(y)\}|\phi_0> \to 
u_0(\vec{x}) \bar u_0(\vec{y})\exp[-i{\cal E}_0(x_0-y_0)]\theta(x_0-y_0). 
\end{eqnarray} 
For meson fields we use the free Feynman propagator for a boson field with  
\begin{eqnarray}
i\Delta_{ij}(x-y)=<0|T\{\Phi_i(x)\Phi_j(y)\}|0>=\delta_{ij} 
\int\frac{d^4k}{(2\pi)^4i}\frac{\exp[-ik(x-y)]}{M_\Phi^2-k^2-i\epsilon}.  
\end{eqnarray}
Superscript $"N"$ in Eq. (\ref{Energy_shift}) indicates that the matrix 
elements are projected on the respective nucleon states. The nucleon wave 
function $|N>$ is conventionally set up by the product of the $SU(6)$ 
spin-flavor w.f. and $SU(3)_c$ color w.f. (see details in \cite{Close}), 
where the nonrelativistic single quark spin w.f. is replaced by the 
relativistic ground state solution of Eq. (\ref{Gaussian_Ansatz}). Projection 
of "one-body" diagrams on the nucleon state refers to 
\begin{eqnarray}
\chi_{f^\prime}^\dagger \chi_{s^\prime}^\dagger \, 
I^{f^\prime f} \, J^{s^\prime s} \, \chi_{f} \chi_{s} 
\, \stackrel{Proj.}\longrightarrow \, 
<N|\sum\limits_{i=1}^{3} (I \, J)^{(i)}|N>
\end{eqnarray} 
where the single particle matrix element of the operators $I$ and $J$, acting  
in flavor and spin space, is replaced by the one embedded in the nucleon 
state. For "two-body" diagrams with two independent quark indices $i$ and $j$ 
the projection prescription reads as 
\begin{eqnarray}
\chi_{f^\prime}^\dagger \chi_{s^\prime}^\dagger \, 
I^{f^\prime f}_1 \, J^{s^\prime s}_1 \, \chi_{f} \chi_{s} \, \otimes 
\chi_{k^\prime}^\dagger \chi_{\sigma^\prime}^\dagger \, 
I^{k^\prime k}_2 \, J^{\sigma^\prime \sigma}_2 \, \chi_{k} \chi_{\sigma} 
\, \stackrel{Proj.}\longrightarrow \, 
<N|\sum\limits_{i \, \not\! = \, j}^{3} (I_1 \, J_1)^{(i)} \otimes 
(I_2 \, J_2)^{(j)}|N> . 
\end{eqnarray}

The total nucleon mass is given by $m_N^{r}=m_N^{core} + \Delta m_N$. 
Superscript $"r"$ refers to the renormalized value of the nucleon 
mass at one loop, that is the order of accuracy we are working in. 
The diagrams that contribute to the nucleon mass shift $\Delta m_N$ at one 
loop are shown in Fig.1: meson cloud (Fig.1a) and meson exchange diagrams  
(Fig.1b). The explicit expression for the nucleon mass 
including one-loop corrections is given by 
\begin{eqnarray}\label{Baryon_mass}
&&m_N^r = m_N^{core} + \Delta m_N = 
3 \, ({\cal E}_0 + \gamma\hat{m}) \, + \, 
\sum\limits_{\Phi = \pi, K, \eta} d_N^\Phi \, \Pi (M_\Phi^2) \\
\mbox{with} \,\,\,\,\, & &d_N^\pi=\frac{171}{400}, \,\,\, 
d_N^K=\frac{6}{19}d_N^\pi, \,\,\, d_N^\eta=\frac{1}{57}d_N^\pi , 
\nonumber
\end{eqnarray} 
where $d_N^\Phi$ are the recoupling coefficients defining the partial 
contribution of the $\pi$, $K$ and $\eta$-meson cloud to the mass shift of 
the nucleon. For the following it is also useful to separately indicate the 
contributions to $d_N^\Phi$ from the meson cloud ($d_N^{\Phi; MC}$) and the 
meson exchange diagrams ($d_N^{\Phi; EX}$):
\begin{eqnarray}\label{recoupling_coeff}
& &d_N^{\pi; MC} = \frac{81}{400}, \hspace*{.5cm} 
d_N^{K; MC}   \equiv d_N^K = \frac{54}{400}, \hspace*{.5cm} 
d_N^{\eta; MC} = \frac{9}{400} \\
\mbox{and}\,\,\,\,& &d_N^{\pi; EX} = \frac{90}{400}, \hspace*{.5cm} 
d_N^{K; EX}    = 0, \hspace*{2.15cm} 
d_N^{\eta; EX} = - \frac{6}{400} . \nonumber
\end{eqnarray}
The self-energy operators $\Pi (M_\Phi^2)$, corresponding to meson cloud 
contributions with definite flavor, differ only in their value for the meson 
mass and are given by 
\begin{eqnarray}\label{Sigma_Phi}
\Pi (M_\Phi^2) \, = \, - \, \biggl(\frac{g_A}{\pi F}\biggr)^2 \,\, 
\int\limits_0^\infty \frac{dp \, p^4}{w_\Phi^2(p^2)} \,\, F_{\pi NN}^2(p^2) .  
\end{eqnarray} 
For a meson with three-momentum $\vec{p}$ the meson energy is   
$w_\Phi(p^2)=\sqrt{M_\Phi^2+p^2}$ with $p=|\vec{p}|$ and $F_{\pi NN}(p^2)$ is 
the $\pi NN$ form factor normalized to unity at zero recoil $(p^2=0)$:  
\begin{eqnarray}\label{F_piNN}
F_{\pi NN}(p^2) = \exp\biggl(-\frac{p^2R^2}{4}\biggr) \biggl\{ 1 \, + \, 
\frac{p^2R^2}{8} \biggl(1 \, - \, \frac{5}{3g_A}\biggr)\biggr\} .  
\end{eqnarray}
Finally, the effect of a finite current quark mass $\hat{m}$ on the nucleon 
mass shift is taken into account perturbatively (for details see \cite{PCQM}), 
resulting in the linear term $3\gamma \hat{m}$ in Eq. (\ref{Baryon_mass}). 

\subsection{Renormalization of the PCQM} 

To redefine our perturbation series up to a given order in terms of 
renormalized quantities a set of counterterms $\delta{\cal L}$ has to be 
introduced in the Lagrangian. Thereby, the counterterms play a dual role: 
i) to maintain the proper definition of physical parameters, such as nucleon 
mass and, in particular, the nucleon charge and ii) to effectively reduce the 
number of Feynman diagrams to be evaluated. 

\subsubsection{Renormalization of the quark field} 

First, we introduce the renormalized quark field $\psi^r$ with renormalized 
mass ${\cal M}^r$, substituting the original field $\psi$. Again, we restrict 
the expansion of the renormalized quark field to the ground state with 
\begin{eqnarray}\label{psi_phys}
\psi^r_i(x; m^r_i)=b_0 u_0^r(\vec{x}; m^r_i) \exp(-i {\cal E}_0^r(m^r_i) t) , 
\end{eqnarray}
where $"i"$ is the SU(3) flavor index; ${\cal E}_0^r(m^r_i)$ is the 
renormalized energy of the quark field in the ground state obtained from 
the solution of the Dirac equation 
\begin{eqnarray}\label{Dirac_Eq_new} 
[- i \vec{\alpha}\vec{\nabla} + \beta m^r_i + \beta S(r)+V(r)-
{\cal E}_0^r(m^r)] u_0^r(\vec{x}; m^r)=0 . 
\end{eqnarray}    
Using the derivations of the previous chapter, the renormalized mass 
$m^r_i$ of the quark field is given by 
\begin{eqnarray}\label{m_r}
& &m^r_u \, = \, m^r_d = \, \hat m^r \, = \,\hat m \, - \delta\hat m  \, 
= \, \hat m \, + \, \frac{1}{3\gamma} \, 
\sum\limits_{\Phi = \pi, K, \eta} d_N^{\Phi; MC} \, \Pi (M_\Phi^2) , \\
& &m_s^r = m_s - \delta m_s = m_s + \, \frac{2}{3\gamma} \, 
\biggl[d_N^{K; MC} \, \Pi (M_K^2) + 2 d_N^{\eta; MC} \, 
\Pi (M_\eta^2)\biggr] , \nonumber 
\end{eqnarray} 
The meson exchange contribution will be included when introducing 
nucleon mass renormalization. For the quark masses we will use in the 
following the compact notation: 
\begin{eqnarray}\label{current_masses}
{\cal M}^r = \mbox{diag} \{ \hat{m}^r, \hat{m}^r, m_s^r \} \hspace*{.5cm}  
\mbox{and} \hspace*{.5cm}  
\delta {\cal M} = \mbox{diag}\{\delta \hat m, \delta \hat m, \delta m_s\}. 
\end{eqnarray}  
The solutions of Eq. (\ref{Dirac_Eq_new}), ${\cal E}_0^r(m^r_i)$ and 
$u_0^r(\vec{x}; m^r_i)$, are functions of $m^r_i$. Obviously, the difference 
between nonstrange and strange quark solutions is solely due to the flavor 
dependent quark mass $m_i^r$. In the limit $m^r_i\to 0$ the solutions for 
nonstrange and strange quarks are degenerate: 
${\cal E}_0^r(\hat m^r) \equiv {\cal E}_0$ and 
$u_0^r(\vec{x}; 0) \equiv u_0(\vec{x})$ (see Eq. (\ref{Dirac_Eq})). For the 
renormalized wave function $u_0^r(\vec{x}; m^r_i)$ we again consider the 
Gaussian ansatz 
\begin{eqnarray}\label{Gaussian_Ansatz_r} 
u_0^r(\vec{x}; m^r_i) \, = \, N(m^r_i) \, 
\exp\biggl[ - c(m^r_i) \frac{\vec{x}^{\, 2}}{2R^2}\biggr] \, 
\left(
\begin{array}{c}
1\\
i \rho(m^r_i) \, \frac{\displaystyle{\vec{\sigma}\vec{x}}}{\displaystyle{R}}\\
\end{array}
\right) \, \chi_s \, \chi_f\, \chi_c  
\end{eqnarray}      
with normalization 
\begin{eqnarray}
\int d^3x \, u^{\dagger r}_0(x; m^r_i) \, u_0^r(x; m^r_i) \equiv 1 . 
\end{eqnarray} 
In Eq. (\ref{Gaussian_Ansatz_r}) the functions $N(m^r_i)$, $c(m^r_i)$ 
and $\rho(m^r_i)$ are normalized at the point $m^r_i = 0$ as follows: 
\begin{eqnarray}
N(0) = N, \hspace*{1cm} c(0) =1, \hspace*{1cm} \rho(0) = \rho
\end{eqnarray} 
The product $\rho(m^r_i) c(m^r_i)$ can be shown to be $m^r_i$-invariant and 
we therefore obtain the additional condition  
\begin{eqnarray}
\rho(m^r_i) \,\, c(m^r_i) \equiv \rho . 
\end{eqnarray}
Treating $m^r_i$ as a small perturbation, Eq. (\ref{Dirac_Eq_new}) 
can be solved perturbatively, resulting in:   
\begin{eqnarray}\label{corr_E_u} 
{\cal E}_0^r(m^r_i) = {\cal E}_0 + \delta {\cal E}_0(m^r_i) 
\,\,\,\,\, \mbox{and} \,\,\,\,\, 
u_0^r(\vec{x}; m^r_i) &=& u_0(\vec{x}) + \delta u_0(\vec{x}; m^r_i) 
\end{eqnarray}
where 
\begin{eqnarray}
\delta {\cal E}_0(m^r_i) = \gamma m^r_i \,\,\,\,\, \mbox{and} \,\,\,\,\, 
\delta u_0^r(\vec{x}; m^r_i) = 
\displaystyle{\frac{m^r_i}{2} \frac{\rho R}{1+\frac{3}{2}\rho^2} 
\biggl( \frac{\frac{1}{2}+\frac{21}{4}\rho^2}
{1+\frac{3}{2}\rho^2} - \frac{\vec{x}^{\, 2}}{R^2} + \gamma^0 
\biggr)} u_0(\vec{x}) . \nonumber  
\end{eqnarray} 
For our set of model parameters the ground state quark energy ${\cal E}_0$ 
is about 400 MeV and for the energy corrections $\delta{\cal E}_0$ relative 
to ${\cal E}_0$ we obtain 
\begin{eqnarray}\label{Energy_PT} 
\bigg|\frac{\delta {\cal E}_0(\hat{m}^r)}{{\cal E}_0}\bigg| \approx 14 \% 
\,\,\,\,\, \mbox{and} \,\,\,\,\, 
\bigg|\frac{\delta {\cal E}_0(m^r_s)}{{\cal E}_0}\bigg| \approx 18 \% . 
\end{eqnarray}
Given the small corrections expressed in Eq. (\ref{Energy_PT}), 
the perturbative treatment of a finite (renormalized) quark mass is 
a meaningful procedure. 

\subsubsection{Renormalized effective Lagrangian} 

Having set up renormalized fields and masses for the quarks we are in the 
position to rewrite the original Lagrangian. The renormalized effective 
Lagrangian including the photon field $A_\mu$ is now written as 
\begin{eqnarray}\label{L_full_r} 
{\cal L}^r_{full} \,= \, {\cal L}_{\psi}^r \, + \, {\cal L}_{\Phi} \, + \, 
{\cal L}_{ph} \, + \, {\cal L}^r_{int} .  
\end{eqnarray} 
The renormalized quark Lagrangian ${\cal L}^r_{\psi}$ defines free nucleon 
dynamics at one-loop with 
\begin{eqnarray}\label{L_psi_r}
& &{\cal L}^r_{\psi} = {\cal L}^r_{\bar\psi\psi} + 
                     {\cal L}^r_{(\bar\psi\psi)^2}, \hspace*{1.5cm} 
{\cal L}^r_{\bar\psi\psi} = 
\bar\psi^r(x)[i\not\!\partial - {\cal M}^r - S(r) - \gamma^0 V(r)]\psi^r(x) \\
& &{\cal L}^r_{(\bar\psi\psi)^2} = c_\pi \sum\limits_{i=1}^{3} 
[\bar\psi^r(x) i\gamma^5 \lambda_i\psi^r(x)]^2  
+ c_K \sum\limits_{i=4}^{7} 
[\bar\psi^r(x) i\gamma^5 \lambda_i\psi^r(x)]^2 
+ c_\eta [\bar\psi^r(x) i\gamma^5 \lambda_8\psi^r(x)]^2 . 
\nonumber
\end{eqnarray} 
The parameters $\delta {\cal M}$ of Eq. (\ref{current_masses}) guarantee the 
proper nucleon mass renormalization due to the meson cloud diagrams of Fig.1a. 
The terms contained in ${\cal L}^r_{(\bar\psi\psi)^2}$ are introduced for the 
purpose of nucleon mass renormalization due to the meson exchange diagram 
of Fig.1b. The corresponding renormalization parameters $c_\pi$, $c_K$ 
and $c_\eta$ are deduced from Eqs. (\ref{Baryon_mass}) and 
(\ref{recoupling_coeff}) as 
\begin{eqnarray}
c_\Phi \, = \, - \frac{9}{200} \, \frac{(2\pi R^2)^{3/2}}{(1-\gamma^2)} \, 
\Pi (M_\Phi^2) . 
\end{eqnarray} 
The free meson Lagrangian ${\cal L}_\Phi$ is written as 
\begin{eqnarray}\label{L_meson}
{\cal L}_\Phi = - \frac{1}{2} \sum\limits_{i,j=1}^{8} 
\Phi_i(x) ( \delta_{ij} \Box + M_{ij}^2 ) \Phi_j(x) 
\end{eqnarray}
where $\Box = \partial^\mu \partial_\mu$ and $M_{ij}^2$ is 
the diagonal meson mass matrix with 
\begin{eqnarray}
M_{11}^2 = M_{22}^2 = M_{33}^2 = M_\pi^2, \hspace*{.75cm}
M_{44}^2 = M_{55}^2 = M_{66}^2 = M_{77}^2 = M_K^2, \hspace*{.75cm}
M_{88}^2 = M_\eta^2 .
\end{eqnarray} 
For the photon field $A_\mu$ we have the usual kinetic term 
\begin{eqnarray}\label{L_ph} 
{\cal L}_{ph} = - \frac{1}{4} F_{\mu\nu}(x)F^{\mu\nu}(x) 
\hspace*{.5cm} \mbox{with} 
\hspace*{.5cm} F_{\mu\nu}(x) = \partial_\nu A_\mu(x) - \partial_\mu A_\nu(x) . 
\end{eqnarray}
The renormalized interaction Lagrangian 
${\cal L}^r_{int} = {\cal L}^r_{str} + {\cal L}^r_{em}$ 
contains a part due to the strong 
\begin{eqnarray}
{\cal L}_{str}^r = {\cal L}^{str}_I + \delta {\cal L}^{str} 
\end{eqnarray}
and the electromagnetic interaction  
\begin{eqnarray}
{\cal L}_{em}^r = {\cal L}^{em}_I + \delta {\cal L}^{em} . 
\end{eqnarray} 
The strong interaction term ${\cal L}^{str}_I$ is given by 
\begin{eqnarray}\label{L_str_I} 
{\cal L}^{str}_I = - \bar\psi^r(x) i\gamma^5 \frac{\hat\Phi(x)}{F} 
S(r) \psi^r(x) 
\end{eqnarray}
The interaction of mesons and quarks with the electromagnetic field is 
described by   
\begin{eqnarray}\label{L_em_I} 
{\cal L}^{em}_I &=& - e A_\mu(x) \bar\psi^r(x) \gamma^\mu Q \psi^r(x) 
- e A_\mu(x) \sum\limits_{i,j=1}^{8}\biggl[f_{3ij} + 
\frac{f_{8ij}}{\sqrt{3}}\biggr] 
\Phi_i(x) \partial^\mu \Phi_j(x)\nonumber\\
 &+& \frac{e^2}{2} A_\mu^2(x) \sum\limits_{i=1,2,4,5} \Phi^2_i(x) . 
\end{eqnarray}
The term ${\cal L}^{em}_I$ is generated by minimal substitution with 
\begin{eqnarray}
& &\partial_\mu\psi^r \to D_\mu\psi^r = \partial_\mu\psi^r + 
i e Q A_\mu \psi^r , \\ 
& &\partial_\mu\Phi_i \to D_\mu\Phi_i = \partial_\mu\Phi_i + 
e \biggl[f_{3ij} + \frac{f_{8ij}}{\sqrt{3}}\biggr] A_\mu \Phi_j 
\nonumber
\end{eqnarray}
where $Q$ is the quark charge matrix and $f_{ijk}$ are the totally 
antisymmetric structure constants of $SU(3)$. 

The set of counterterms, denoted by $\delta {\cal L}^{str}$ and 
$\delta {\cal L}^{em}$, is explicitly given by 
\begin{eqnarray}\label{counter_terms}
\delta {\cal L}^{str}&=& \delta {\cal L}^{str}_1 + \delta {\cal L}^{str}_2 + 
\delta {\cal L}^{str}_3, \label{L_ct_all}\nonumber\\
\mbox{with}& &\nonumber\\ 
\delta {\cal L}^{str}_1&=& \bar\psi^r(x) \, (Z - 1) \, [i\not\! \partial 
- {\cal M}^r - S(r) - \gamma^0V(r)]\psi^r(x), \label{L_ct_str1}\nonumber\\
& &\nonumber\\  
\delta {\cal L}^{str}_2&=& - \, \bar\psi^r(x) \, \delta {\cal M} \, \psi^r(x), 
\label{L_ct_str2}\nonumber\\[3mm] 
\delta {\cal L}^{str}_3&=& - c_\pi \sum\limits_{i=1}^{3} 
[\bar\psi^r(x) i\gamma^5 \lambda_i\psi^r(x)]^2  - c_K \sum\limits_{i=4}^{7} 
[\bar\psi^r(x) i\gamma^5 \lambda_i\psi^r(x)]^2 
- c_\eta [\bar\psi^r(x) i\gamma^5 \lambda_8\psi^r(x)]^2, \label{L_ct_str4}
\nonumber\\
\mbox{and}& &\nonumber\\ 
\delta {\cal L}^{em}&=& - e A_\mu(x) \bar\psi^r(x) \, (Z - 1) \, 
\gamma^\mu Q \psi^r(x) . \label{L_ct_em} \nonumber  
\end{eqnarray}
Here, $Z = {\rm diag} \{\hat Z, \hat Z, Z_s \}$ is the diagonal matrix of  
renormalization constants ($\hat Z$ for $u,d$-quarks and $Z_s$ for 
$s$-quark). The values of $\hat Z$ and $Z_s$ are determined by the charge 
conservation condition. The simplest way to fix $\hat Z$ and $Z_s$ is on the 
quark level. The same set of values for $\hat Z$ and $Z_s$ is also obtained 
when requiring charge conservation on the baryon level.  
Results for $\hat{Z}$ and $Z_s$ will be discussed below. 

Now we briefly explain the role of each counterterm and why the set of 
constants $\hat Z$ and $Z_s$ is identical in  $\delta {\cal L}^{str}_1$  
and $\delta {\cal L}^{em}$. The counterterm $\delta {\cal L}^{em}$ is 
introduced to guarantee charge conservation. The counterterm 
$\delta {\cal L}^{str}_1$, containing the same renormalization 
constants $\hat Z$ and $Z_s$ as in $\delta {\cal L}^{em}$, is added to fulfil 
electromagnetic local gauge invariance on the Lagrangian level. The same term 
also leads to conservation of the vector current (baryon number conservation). 
Alternatively, $\delta {\cal L}^{em}$ can also be deduced from 
$\delta {\cal L}^{str}_1$ by minimal substitution. In covariant theories 
the equality of the renormalization constants in $\delta {\cal L}^{str}_1$ 
and $\delta {\cal L}^{em}$ is known as the Ward identity. The counterterms 
$\delta {\cal L}^{str}_2$ and $\delta {\cal L}^{str}_3$ compensate the 
contributions of the meson cloud (Fig.1a) and meson exchange diagrams (Fig.1b) 
to the nucleon mass $m_N^r$ (The contribution of meson cloud and exchange 
diagrams is already taken into account in the renormalized quark Lagrangian 
${\cal L}^r_{\psi}$.). 
  
\subsubsection{Renormalization of nucleon mass and charge}

Now we illustrate the explicit role of the counterterms when performing the 
calculation of the nucleon mass and the nucleon charge. The renormalized 
nucleon mass $m_N^r$ is defined by the expectation value of the Hamiltonian 
${\cal H}^r_\psi$ (as derived from the Lagrangian ${\cal L}^r_\psi$)  
averaged over state $|\phi_0>$ and projected on the respective nucleon 
states:  
\begin{eqnarray}\label{Fig2a} 
m_N^r \, \equiv \, ^N\!\!<\phi_0| \, \int \, \delta(t) \, d^4x \,  
{\cal H}^r_\psi(x) \, |\phi_0>^{N},   
\end{eqnarray}
By inclusion of the counterterms the strong interaction Lagrangian 
${\cal L}^{str}_r$ should give a zero contribution to the shift of the 
renormalized nucleon mass at one loop, that is 
\begin{eqnarray}\label{Delta_m_N^r_new}
\Delta m_N^r\, &=& \, ^N\!\!<\phi_0| \, \sum\limits_{n=1}^{2} \frac{i^n}{n!} 
\, \int \, i\delta(t_1) \, d^4x_1 \, \ldots \, d^4x_n \,\,  
T[{\cal L}^{str}_r(x_1) \,\ldots \,{\cal L}^{str}_r(x_n)] \, |\phi_0>_{c}^{N}\\
&=& \, ^N\!\!<\phi_0| \, -\frac{i}{2} \int \, \delta(t_1) \, d^4x_1  \,  
d^4x_2 \,\,  T[{\cal L}^{str}_I(x_1) {\cal L}^{str}_I(x_2)] \, |\phi_0>_{c}^{N}
\nonumber\\
&-& \, ^N\!\!<\phi_0| \, \int \, \delta(t) \, d^4x \,\, \sum\limits_{i=1}^3  
\delta {\cal L}^{str}_i(x)\,|\phi_0>^{N} \equiv 0 . \nonumber
\end{eqnarray}
The propagator of the renormalized quark field $\psi^r$ is given by 
\begin{eqnarray} 
\hspace*{-.75cm} 
iG_{\psi^r}(x,y)=<\phi_0|T\{\psi^r(x)\bar\psi^r(y)\}|\phi_0> \to 
u_0^r(\vec{x}) \bar u_0^r(\vec{y})\exp[-i{\cal E}_0^r(x_0-y_0)]
\theta(x_0-y_0) . 
\end{eqnarray} 
It differs from the unperturbed quark propagator $iG_{\psi}(x,y)$ 
by terms of order $\hat m^r$, which in turn only contribute to the two-loop 
calculations. Thus, to the order of accuracy we are working in (up to one-loop 
perturbation theory) it is sufficient to use the unperturbed quark propagator 
$iG_{\psi}(x,y)$ instead of the renormalized one. 

To prove  Eq. (\ref{Delta_m_N^r_new}), we first note that the contribution of 
the counterterm $\delta {\cal L}^{str}_1$ is equal to zero due to the 
equation of motion (\ref{Dirac_Eq_new}), that is 
\begin{eqnarray}
^N\!\!<\phi_0|\, \int \, \delta(t) \, d^4x \, \delta {\cal L}^{str}_1(x) \, 
|\phi_0>^{N}\equiv 0 . 
\end{eqnarray}
The counterterms $\delta {\cal L}^{str}_2$ and $\delta {\cal L}^{str}_3$ 
compensate the contribution of the meson cloud (Fig.1a) and exchange diagrams  
(Fig.1b), respectively, with 
\begin{eqnarray}\label{Delta_m_N^r_new_next} 
& &^N\!\!<\phi_0| \, -\frac{i}{2} \,\int \, \delta(t_1) \, d^4x_1 \, d^4x_2 
\, \,  T[{\cal L}^{str}_I(x_1) {\cal L}^{str}_I(x_2)] \, |\phi_0>_{c}^{N} \\
&-&^N\!\!<\phi_0|\,\int \, \delta(t) \, d^4x \, [\delta {\cal L}^{str}_2(x) \, 
+ \delta {\cal L}^{str}_3(x)] \, |\phi_0>^{N} \equiv 0 , \nonumber 
\end{eqnarray}
hence Eq. (\ref{Delta_m_N^r_new}) is fulfilled. The calculation of the nucleon 
mass $m_N^r$ at one-loop can then either be done with the "unrenormalized" 
Lagrangian ${\cal L}_{eff}$ or with the "renormalized" 
version ${\cal L}_{full}^r$ (\ref{L_full_r}). Both results for $m_N^r$ are 
identical and are given by Eq. (\ref{Baryon_mass}). 

Now we consider the nucleon charge and prove that the properly introduced 
counterterms guarantee charge conservation. Using Noether's theorem we first 
derive from the renormalized Lagrangian (\ref{L_full_r}) the electromagnetic,  
renormalized current operator: 
\begin{eqnarray}\label{em_current} 
j^\mu_r = j^\mu_{\psi^r} \, + \, j^\mu_\Phi \, + \delta j^\mu_{\psi^r} .  
\end{eqnarray}
It contains the quark component $j^\mu_{\psi^r}$, the charged meson component 
$j^\mu_\Phi$ and the contribution of the counterterm $\delta j^\mu_{\psi^r}$:  
\begin{eqnarray}
j^\mu_{\psi^r} &=& \bar\psi^r \gamma^\mu Q \psi^r 
\equiv \frac{1}{3} \,\,  
[ 2 \, \bar u^r \gamma^\mu u^r - \bar d^r \gamma^\mu d^r  
- \bar s^r \gamma^\mu s^r] ,\label{em_psi_currents}\\
j^\mu_\Phi &=& \biggl[f_{3ij} + \frac{f_{8ij}}{\sqrt{3}}\biggr] \Phi_i 
\partial^\mu \Phi_j \equiv \,\, [\pi^- i\partial^\mu \pi^+ - 
\pi^+ i\partial^\mu \pi^- \, + \, K^- i\partial^\mu K^+ - 
K^+ i\partial^\mu K^-],\label{em_psi_Phi_currents}\nonumber\\
\mbox{and}& &\nonumber\\
\delta j^\mu_{\psi^r} &=&  \bar\psi^r \, (Z - 1) \, \gamma^\mu Q \psi^r 
\equiv \frac{1}{3} \,   
[ 2 \, (\hat{Z} - 1) \, \bar u^r \gamma^\mu u^r - (\hat{Z} - 1) \, 
\bar d^r \gamma^\mu d^r  - (Z_s - 1) \bar s^r \gamma^\mu s^r] . 
\label{em_ct_currents} \nonumber        
\end{eqnarray}
The renormalized nucleon charge $Q_N^r$ at one loop is defined as 
\begin{eqnarray}\label{Q_Nr2}
Q_N^r \, = \, ^N\!\!<\phi_0| \, \sum\limits_{n=0}^{2} \, \frac{i^{n}}{n!} \,  
\int \, \delta(t) \, d^4x \, d^4x_1 \ldots d^4x_n \, 
T[{\cal L}^{str}_r(x_1) \, \ldots \, {\cal L}^{str}_r(x_n)\, j^0_r(x)]\, 
|\phi_0>_c^{N} . 
\end{eqnarray}
Charge conservation requires that the nucleon charge is not changed after 
renormalization, that is 
\begin{eqnarray}\label{Q_Nr-Q_N}
Q_N^r \equiv Q_N = \left\{
\begin{array}{ll}
1 & \hspace*{.5cm} \mbox{for} \hspace*{.5cm} 
                          N=p \hspace*{.5cm} \mbox{(proton)}    \\
0 & \hspace*{.5cm} \mbox{for} \hspace*{.5cm} 
                          N=n \hspace*{.5cm} \mbox{(neutron)} . \\
\end{array}
\right.    
\end{eqnarray}
Thereby, $Q_N$ is the nucleon charge in the three-quark core approximation,  
which is defined as the expectation value of the quark charge operator 
$\hat{Q}_\psi \, = \, \int d^3x \, j^0_{\psi}(x)$ taken between the 
unperturbed 3q-states $|\phi_0>$: 
\begin{eqnarray}\label{Q_N}
Q_N \, = \, ^N\!\!<\phi_0| \, \int \, \delta(t) d^4x \, j^0_{\psi}(x) \, 
|\phi_0>^{N} . 
\end{eqnarray}
Eqs. (\ref{Q_Nr2})-(\ref{Q_N}) completely define {\it the charge 
conservation condition} within our approach.  

From nucleon charge conservation at one loop we obtain a condition on the 
renormalization constant $\hat Z$. To fix the constant $Z_s$ we should 
consider the charge conservation of baryons containing $s$-quarks, e.g. 
$\Sigma^+$ - baryon. In the  one-loop approximation following diagrams 
contribute to the nucleon charge $Q_N^r$ (see Figs.2a-l): three-quark diagram 
(Fig.2a) with insertion of the quark current $j^\mu_{\psi^r}$, three-quark 
diagram (Fig.2b) with the counterterm $\delta j^\mu_{\psi^r}$ (three-quark 
counterterm diagram), meson-cloud diagram (Fig.2c) with the meson current 
$j^\mu_\Phi$, vertex correction diagram (Fig.2d) with the quark current 
$j^\mu_{\psi^r}$, self-energy diagrams (Figs.2e and 2f) and exchange current 
diagrams (Figs.2i and 2j) with insertion of the quark current 
$j^\mu_{\psi^r}$. We also obtain a set of diagrams (Figs.2g, 2h, 2k and 2l) 
generated by the counterterms $\delta {\cal L}^{str}_2(x)$ and 
$\delta {\cal L}^{str}_3(x)$. The contribution of the counterterm 
$\delta {\cal L}^{str}_1(x)$ is equal to zero due to the equation of 
motion (\ref{Dirac_Eq_new}). By definition of the counterterms 
$\delta {\cal L}^{str}_2(x)$ and $\delta {\cal L}^{str}_3(x)$, the self-energy 
and the meson exchange current diagrams of Fig.2e, f, i, j are compensated by 
the counterterm diagrams of Fig.2g, h, k and l, respectively. 

The contribution of the three-quark diagram (Fig.2a) to the nucleon charge 
is trivially given by 
\begin{eqnarray}
Q_N^{r; a} \, = \, ^N\!\!<\phi_0|\, \int \, \delta(t) d^4x \, j^0_{\psi^r}(x) 
\, |\phi_0>^{N}  \equiv Q_N . 
\end{eqnarray}
The three-quark counterterm diagram (Fig.2b) is simply related to the one of 
Fig.2a with: 
\begin{eqnarray}
Q_N^{r; b} \, = \, (\hat{Z} - 1) \, Q_N^{r; a} \, = \, (\hat{Z} - 1) \, 
^N\!\!<\phi_0|\, \int \, \delta(t) d^4x \, j^0_{\psi^r}(x) \, |\phi_0>^{N} 
\equiv (\hat{Z} - 1) \, Q_N . 
\end{eqnarray}
The meson cloud diagram (Fig.2c) generates the term 
\begin{eqnarray}
Q_N^{r; c} = \frac{27}{400} \biggl(\frac{g_A}{\pi F}\biggr)^2 
\int\limits_0^\infty dp \, p^4 \,\, F_{\pi NN}^2(p^2) 
\sum\limits_{\Phi = \pi, K} \frac{q_N^{\Phi; c}}{w_\Phi^3(p^2)}
\end{eqnarray}
where 
\begin{eqnarray}\label{matrices_1}
q_N^{\pi; c} = \left\{
\begin{array}{rl}
  \frac{\displaystyle{2}}{\displaystyle{3}} & \mbox{for} \,\, N=p \\
  &\\
- \frac{\displaystyle{2}}{\displaystyle{3}} & \mbox{for} \,\, N=n \\
\end{array}
\right., \hspace*{1cm}
q_N^{K; c} = \left\{
\begin{array}{rl}
  \frac{\displaystyle{4}}{\displaystyle{3}} & \mbox{for} \,\, N=p \\
  &\\
  \frac{\displaystyle{2}}{\displaystyle{3}} & \mbox{for} \,\, N=n \\
\end{array}
\right. . \nonumber 
\end{eqnarray}
The contribution of the vertex correction diagram (Fig.2d) is given by 
\begin{eqnarray}
Q_N^{r; d} = \frac{27}{400} \biggl(\frac{g_A}{\pi F}\biggr)^2 
\int\limits_0^\infty dp \, p^4 \,\, F_{\pi NN}^2(p^2) 
\sum\limits_{\Phi = \pi, K, \eta} 
\frac{q_N^{\Phi; d}}{w_\Phi^3(p^2)}
\end{eqnarray}
where 
\begin{eqnarray}\label{matrices_2}
q_N^{\pi; d} = \left\{
\begin{array}{rl}
  \frac{\displaystyle{1}}{\displaystyle{3}} & \mbox{for} \,\, N=p \\
  &\\
  \frac{\displaystyle{2}}{\displaystyle{3}} & \mbox{for} \,\, N=n \\
\end{array}
\right., 
\hspace*{1cm}
q_N^{K; d} = \left\{
\begin{array}{rl}
 - \frac{\displaystyle{2}}{\displaystyle{3}} & \mbox{for} \,\, N=p \\
 &\\
 - \frac{\displaystyle{2}}{\displaystyle{3}} & \mbox{for} \,\, N=n \\
\end{array}
\right., 
\hspace*{1cm}
q_N^{\eta; d} = \left\{
\begin{array}{rl}
   \frac{\displaystyle{1}}{\displaystyle{9}} & \mbox{for} \,\, N=p \\
   &\\
   0                                         & \mbox{for} \,\, N=n \\
\end{array}
\right. . \nonumber 
\end{eqnarray}
To guarantee charge conservation, the sum of meson-cloud and vertex correction 
diagrams must be compensated by the counterterm contribution:  
\begin{eqnarray}
Q_N^{r; b} + Q_N^{r; c} + Q_N^{r; d} \equiv 0 . 
\end{eqnarray}
The last requirement fixes the value of the renormalization constant 
$\hat Z$ at one loop to 
\begin{eqnarray}\label{Z-eq}
\hspace*{-.7cm} 
\hat Z = 1 - \, \frac{27}{400} \biggl(\frac{g_A}{\pi F}\biggr)^2 
\int\limits_0^\infty dp \, p^4 \,\, F_{\pi NN}^2(p^2) 
\biggl\{ \frac{1}{w_\pi^3(p^2)} + \frac{2}{3w_K^3(p^2)} 
+ \frac{1}{9w_\eta^3(p^2)}  \biggr\} . 
\end{eqnarray}
In the two-flavor picture, that is when we restrict to the pion cloud 
contribution only, we obtain a value of $\hat Z =0.9 \pm 0.02$ for our set of 
parameters. The contribution of kaon and $\eta$-meson loops to the constant 
$\hat Z$ is strongly suppressed due to the energy denominators in 
Eq. (\ref{Z-eq}). In the three-flavor picture we get $\hat Z=0.88 \pm 0.03$, 
which deviates only slightly from the two-flavor result. The minor role of 
kaon and $\eta$-meson loop contributions to nucleon properties was also found 
in our previous analysis of meson-nucleon sigma-terms \cite{PCQM}. As already 
mentioned, the renormalization constant $Z_s$ is fixed from the charge 
conservation of baryons containing strange quarks (e.g. $\Sigma^+$-baryon). 
Here we obtain the analytical result:  
\begin{eqnarray}\label{Zs-eq} 
Z_s = 1 - \, \frac{27}{400} \biggl(\frac{g_A}{\pi F}\biggr)^2 
\int\limits_0^\infty dp \, p^4 \,\, F_{\pi NN}^2(p^2) 
\biggl\{ \frac{4}{3w_K^3(p^2)} 
+ \frac{4}{9w_\eta^3(p^2)}  \biggr\} . 
\end{eqnarray}
In the $SU(3)$ flavor symmetry limit $(m_u = m_d = m_s)$ both renormalization 
constants $\hat Z$ and $Z_s$ are degenerate. Again, charge conservation within 
our approach is fulfilled both on the quark level (when we directly calculate 
the charge of $u$, $d$ or $s$-quark at one loop) and on the baryon level.  
With the value of $\hat{Z}$ being close to unity for our set of parameters 
the perturbative treatment of the meson cloud is also justified. 

\section{Electromagnetic Form Factors of the Nucleon} 

It is well-known that noncovariant hadron models should explicitly address 
the question of local gauge invariance of electromagnetic form factors 
(for a detailed discussion see Refs. 
\cite{Ernst-Sachs-Wali}-\cite{Ivanov_Locher_Lyubovitskij}). Particularly, 
even when starting from a gauge-invariant Lagrangian the resulting physical 
amplitudes are not necessarily gauge invariant in an arbitrary frame due to 
the use of noncovariant techniques (e.g. noncovariant particle propagators).  
The lack of Lorentz covariance in our approach is linked to the static 
potential solutions for the quark fields and, as consequence, to the use of 
a noncovariant quark propagator. However, when restricting our kinematics to 
a specific frame, that is the Breit frame, gauge invariance is fulfilled due 
to the decoupling of the time and vector components of the electromagnetic 
current operator. 

Next we follow the original work by Sachs et al. \cite{Ernst-Sachs-Wali,Sachs} 
and Thomas et al. \cite{Theberge-Thomas,Miller-Thomas,Lu-Thomas-Williams} to 
set up the formalism for the electromagnetic form factors of the nucleon. In 
the Breit frame the initial momentum of the nucleon is $p = (E, -\vec{q}/2)$, 
the final momentum is $p^\prime = (E, \vec{q}/2)$ and the 4-momentum of the 
photon is $q = (0, \vec{q}\,)$ with $p^\prime = p + q$. With the space-like 
momentum transfer squared given as $Q^2 = - q^2 = \vec{q}^{\, 2}$, in the 
Breit frame the nucleon charge $G_E^N$ and magnetic $G_M^N$ (Sachs) form 
factors are defined by \cite{Lu-Thomas-Williams} 
\begin{eqnarray}
& &<N_{s^\prime}\biggl(\frac{\vec{q}}{2}\biggr)|J^0(0)|
N_s\biggl(-\frac{\vec{q}}{2}\biggr)> 
= \chi^\dagger_{N_{s^\prime}} \chi_{N_s} G_E^N(Q^2) , \\ 
& &<N_{s^\prime}\biggl(\frac{\vec{q}}{2}\biggr)|\vec{J}(0)|
N_s\biggl(-\frac{\vec{q}}{2}\biggr)> 
= \chi^\dagger_{N_{s^\prime}} \frac{i \vec{\sigma}_N \times \vec{q}}{2m_N} 
\chi_{N_s} G_M^N(Q^2) . \nonumber  
\end{eqnarray}
Here, $J^0(0)$ and $\vec{J}(0)$ are the time and space component of the 
electromagnetic current operator (\ref{em_current}); $\chi_{N_s}$ and 
$\chi^\dagger_{N_{s^\prime}}$ are the nucleon spin w.f. in the initial and 
final state; $\vec{\sigma}_N$ is the nucleon spin matrix. Electromagnetic 
gauge invariance on the baryon level is fulfilled in the Breit frame. In 
particular, for on-mass shell nucleons we verify 
the Ward-Takahashi identity without referring to the explicit form of the 
nucleon charge and magnetic form factors: 
\begin{eqnarray}
q_\mu <N_{s^\prime}\biggl(\frac{\vec{q}}{2}\biggr)|j^\mu_r(0)|
N_s\biggl(-\frac{\vec{q}}{2}\biggr)> &=& \underbrace{q_0}_{\equiv 0} \cdot 
<N_{s^\prime}\biggl(\frac{\vec{q}}{2}\biggr)| j_r^0(0)|
N_s\biggl(-\frac{\vec{q}}{2}\biggr)> \\
+\vec{q} \, \cdot <N_{s^\prime}\biggl(\frac{\vec{q}}{2}\biggr)|\vec{j}_r(0)|
N_s\biggl(-\frac{\vec{q}}{2}\biggr)> &=& \chi^\dagger_{N_{s^\prime}} \, 
\vec{q} \cdot \frac{i \vec{\sigma}_N \times \vec{q}}{2m_N} \, \chi_{N_s}  
\, G_M^N(Q^2) \,\equiv \, 0  
\nonumber 
\end{eqnarray}
At zero recoil ($q^2=0$) the Sachs form factors satisfy the following 
normalization conditions
\begin{eqnarray}
G_E^p(0)=1, \hspace*{1cm} G_E^n(0)=0, \hspace*{1cm} 
G_M^p(0)=\mu_p, \hspace*{1cm} G_M^n(0)=\mu_n, 
\end{eqnarray}
where $\mu_p$ and $\mu_n$ are the magnetic moments of the proton and the 
neutron, respectively. 

The charge radii of the nucleons are given by 
\begin{eqnarray}
& &<r^2>^p_E = - 6 \frac{dG^p_E(Q^2)}{dQ^2}
\Bigg|_{\displaystyle{Q^2 = 0}} , \hspace*{1.5cm}
<r^2>^n_E = -6 \, \frac{dG^n_E(Q^2)}{dQ^2}
\Bigg|_{\displaystyle{Q^2 = 0}} ,\\
& &<r^2>^N_M = - \frac{6}{G^N_M(0)} \frac{dG^N_M(Q^2)}{dQ^2}
\Bigg|_{\displaystyle{Q^2 = 0}} . \nonumber
\end{eqnarray}

In the PCQM the charge and magnetic form factors of nucleon are defined by 
\begin{eqnarray}
\chi^\dagger_{N_{s^\prime}} \chi_{N_s} G_E^N(Q^2)  
&=& ^N\!\!<\phi_0| \, \sum\limits_{n=0}^{2} \frac{i^n}{n!} \, 
\int \, \delta(t) \, d^4x \, d^4x_1 \, \ldots \, d^4x_n \, e^{-iqx} \, 
\label{GEN}\\ 
&\times& T[{\cal L}^{str}_r(x_1) \, \ldots \, {\cal L}^{str}_r(x_n) \, 
j^0_r(x)] \, |\phi_0>_{c}^{N} , \nonumber \\ 
\chi^\dagger_{N_{s^\prime}} \frac{i \vec{\sigma}_N \times \vec{q}}{2m_N} 
\chi_{N_s} G_M^N(Q^2) 
&=& ^N\!\!<\phi_0| \, \sum\limits_{n=0}^{2} \frac{i^n}{n!} \, 
\int \, \delta(t) \, d^4x \, d^4x_1 \, \ldots \, d^4x_n \, 
e^{-iqx} \, \label{GMN} \nonumber \\
&\times& T[{\cal L}^{str}_r(x_1) \, \ldots \, {\cal L}^{str}_r(x_n) \, 
\vec{j}_r(x)] \, |\phi_0>_{c}^{N} . \nonumber 
\end{eqnarray}
The diagrams of Figs.2a-d contribute to the charge form factor of the 
nucleon. For the magnetic form factors we have to include an additional term 
due to the "meson-in-flight" diagram indicated in Fig.3. The other possible 
diagrams at one-loop (like self-energy current and meson-exchange current 
diagrams of Fig.2) are compensated by the counterterms (see previous 
discussion about renormalization of the nucleon charge).   

In the following we indicate the analytical expressions for the relevant 
diagrams: 

a) Tree-quark diagram $(3q)$ (Fig.2a): 
\begin{eqnarray}
G_E^p(Q^2)\bigg|_{3q} &=& G_E^p(Q^2)\bigg|_{3q}^{LO} + 
G_E^p(Q^2)\bigg|_{3q}^{NLO}, 
\hspace*{.5cm} G_E^n(Q^2)\bigg|_{3q}  \equiv 0, \\
G_M^p(Q^2)\bigg|_{3q} &=& G_M^p(Q^2)\bigg|_{3q}^{LO} + 
G_M^p(Q^2)\bigg|_{3q}^{NLO}, \hspace*{.3cm} 
G_M^n(Q^2)\bigg|_{3q} \equiv
- \frac{2}{3} G_M^p(Q^2)\bigg|_{3q},  \nonumber 
\end{eqnarray}
where $G_{E, M}^N(Q^2)\bigg|_{3q}^{LO}$ are the leading-order (LO) terms of 
the tree-quark diagram evaluated with the unperturbed quark w.f. 
$u_0(\vec{x})$; $G_E^p(Q^2)\bigg|_{3q}^{NLO}$ is a correction due to the 
modification of the quark w.f. $u_0(\vec{x}) \to u_0^r(\vec{x};\hat m^r)$ 
referred to as next-to-leading order (NLO): 
\begin{eqnarray}\label{GEP_3q}
G_E^p(Q^2)\bigg|_{3q}^{LO} &=& \exp\biggl(-\frac{Q^2R^2}{4}\biggr) 
\biggl(1 - \frac{\rho^2}{1+\frac{3}{2}\rho^2} \frac{Q^2R^2}{4} \biggr),\\ 
G_E^p(Q^2)\bigg|_{3q}^{NLO} &=& \exp\biggl(-\frac{Q^2R^2}{4}\biggr) 
\, \hat m^r \frac{Q^2 R^3 \rho}{4(1+\frac{3}{2}\rho^2)^2} 
\biggl( \frac{1+7\rho^2+\frac{15}{4}\rho^4}{1+\frac{3}{2}\rho^2} 
- \frac{Q^2 R^2}{4} \rho^2 \biggr), \nonumber \\
G_M^p(Q^2)\bigg|_{3q}^{LO} &=& \exp\biggl(-\frac{Q^2R^2}{4}\biggr) 
\frac{2m_N \rho R}{1+\frac{3}{2}\rho^2}, \nonumber \\ 
G_M^p(Q^2)\bigg|_{3q}^{NLO} &=& G_M^p(Q^2)\bigg|_{3q}^{LO} \, 
\cdot \,\hat m^r \frac{R \rho}{1+\frac{3}{2}\rho^2} \biggl( \frac{Q^2 R^2}{4} 
- \frac{2-\frac{3}{2}\rho^2}{1+\frac{3}{2}\rho^2}\biggr) . \nonumber 
\end{eqnarray}

b) Three-quark counterterm (CT) (Fig.2b):
\begin{eqnarray}
& &G_E^p(Q^2)\bigg|_{CT}\equiv (\hat Z-1) \, G_E^p(Q^2)\bigg|_{3q}^{LO}, 
\hspace*{.7cm} G_E^n(Q^2)\bigg|_{CT} \equiv 0, \\
& &G_M^p(Q^2)\bigg|_{CT}\equiv (\hat Z-1) \, G_M^p(Q^2)\bigg|_{3q}^{LO}, 
\hspace*{.5cm} 
G_M^n(Q^2)\bigg|_{CT}\equiv - \frac{2}{3} G_M^p(Q^2)\bigg|_{CT} . \nonumber 
\end{eqnarray}

c) Meson-cloud diagram (MC) (Fig.2c):
\begin{eqnarray}
\hspace*{-.2cm}
G_E^N(Q^2)\bigg|_{MC}  &=& \frac{9}{400} \, \biggl(\frac{g_A}{\pi F}\biggr)^2 
\int\limits_0^\infty dp \, p^2 \, \int\limits_{-1}^1 dx \, 
(p^2+p\sqrt{Q^2}x) \, {\cal F}_{\pi NN}(p^2,Q^2,x) \, 
t_E^N(p^2,Q^2,x)\bigg|_{MC} \\
\hspace*{-.2cm}G_M^N(Q^2)\bigg|_{MC}  &=& \frac{3}{40} \, m_N \, 
\biggl(\frac{g_A}{\pi F}\biggr)^2 \int\limits_0^\infty dp \, p^4 \,  
\int\limits_{-1}^1 dx (1-x^2) \,\, {\cal F}_{\pi NN}(p^2,Q^2,x) \, 
t_M^N(p^2,Q^2,x)\bigg|_{MC} \nonumber 
\end{eqnarray}
where 
\begin{eqnarray}
{\cal F}_{\pi NN}(p^2,Q^2,x)&=&F_{\pi NN}(p^2) \, 
F_{\pi NN}(p^2+Q^2+2p\sqrt{Q^2}x) , \nonumber\\
t_E^p(p^2,Q^2,x)\bigg|_{MC}&=& 
C_\pi^{11}(p^2,Q^2,x)   + 2 C_K^{11}(p^2,Q^2,x) , \nonumber\\ 
t_E^n(p^2,Q^2,x)\bigg|_{MC}&=& 
- C_\pi^{11}(p^2,Q^2,x) + C_K^{11}(p^2,Q^2,x) , \nonumber\\
t_M^p(p^2,Q^2,x)\bigg|_{MC}&=&
D_\pi^{22}(p^2,Q^2,x)   + \frac{4}{5} D_K^{22}(p^2,Q^2,x), \nonumber\\
t_M^n(p^2,Q^2,x)\bigg|_{MC}&=& 
- D_\pi^{22}(p^2,Q^2,x) - \frac{1}{5} D_K^{22}(p^2,Q^2,x),  \nonumber\\
D^{n_1n_2}_\Phi(p^2,Q^2,x) &=& 
\frac{1}{w_\Phi^{n_1}(p^2)w_\Phi^{n_2}(p^2 + Q^2 + 2p\sqrt{Q^2}x)},\nonumber\\ 
C^{n_1n_2}_\Phi(p^2,Q^2,x) &=& \frac{2D^{n_1n_2}_\Phi(p^2,Q^2,x)} 
{w_\Phi(p^2)+w_\Phi(p^2 + Q^2 + 2p\sqrt{Q^2}x)} . \nonumber 
\end{eqnarray}

d) Vertex-correction  diagram (VC) (Fig.2d):
\begin{eqnarray}
G_{E(M)}^N(Q^2)\bigg|_{VC} = G_{E(M)}^p(Q^2)\bigg|_{3q}^{LO} \, 
\cdot \, \frac{9}{200} \, \biggl(\frac{g_A}{\pi F}\biggr)^2 
\int\limits_0^\infty dp \, p^4 \,\, F_{\pi NN}^2(p^2) 
t_{E(M)}^N(p^2)\bigg|_{VC}, 
\end{eqnarray}
where 
\begin{eqnarray}
t_E^p(p^2)\bigg|_{VC}&=& 
\frac{1}{2} W_\pi(p^2) - W_K(p^2) + \frac{1}{6} W_\eta(p^2) , \nonumber\\ 
t_E^n(p^2)\bigg|_{VC}&=& W_\pi(p^2) - W_K(p^2), \nonumber\\
t_M^p(p^2)\bigg|_{VC}&=& 
\frac{1}{18} W_\pi(p^2) + \frac{1}{9} W_K(p^2) - \frac{1}{18} W_\eta(p^2) , 
\nonumber \\ 
t_M^n(p^2)\bigg|_{VC}&=& 
- \frac{2}{9} W_\pi(p^2) + \frac{1}{9} W_K(p^2) + \frac{1}{27} W_\eta(p^2),
\nonumber \\ 
W_\Phi(p^2) &=& \frac{1}{w_\Phi^3(p^2)} . \nonumber 
\end{eqnarray}

e) Meson-in-flight diagram (MF) (Fig.3):
\begin{eqnarray}
G_E^p(Q^2)\bigg|_{MF}&\equiv& 0, 
\hspace*{.5cm}G_E^n(Q^2)\bigg|_{MF} \, \equiv \, 0, \hspace*{.5cm}
G_M^n(Q^2)\bigg|_{MF} \equiv - G_M^p(Q^2)\bigg|_{MF}\\
G_M^p(Q^2)\bigg|_{MF} &=& \frac{9}{100} \, m_N \, 
\biggl(\frac{g_A}{\pi F}\biggr)^2 \int\limits_0^\infty dp \, p^4 \, 
\int\limits_{-1}^{1} dx (1-x^2) \, {\cal F}_{\pi NN}(p^2,Q^2,x) 
D^{22}_\pi(p^2,Q^2,x) \nonumber . 
\end{eqnarray}
We do not include the modification of the quark w.f. in the calculation of 
one-loop diagrams (Figs.2b, 2c, 2d and 3) since it only gives corrections to 
the next (two-loop) order in perturbation theory. 

We start our analysis with the magnetic moments of nucleons, $\mu_p$ and 
$\mu_n$, given by the expressions  (in units of nucleon magnetons) 
\begin{eqnarray}\label{mu_pmu_n}
\mu_p&=&\mu_p^{LO} \biggl[ 1 \, + \, \delta 
-  \, \frac{1}{400} \, \biggl(\frac{g_A}{\pi F}\biggr)^2 
\int\limits_0^\infty dp \, p^4 \,\, F_{\pi NN}^2(p^2) 
\biggl\{ \frac{26}{w_\pi^3(p^2)} + 
\frac{16}{w_K^3(p^2)} + \frac{4}{w_\eta^3(p^2)} \biggr\}\biggr]\\
&\, + \,& \frac{m_N}{50} \, \biggl(\frac{g_A}{\pi F}\biggr)^2 
\int\limits_0^\infty dp \, p^4 \,\, F_{\pi NN}^2(p^2) 
\biggl\{ \frac{11}{w_\pi^4(p^2)} + \frac{4}{w_K^4(p^2)} \biggr\}\nonumber\\
\mu_n&=& - \frac{2}{3} \, \mu_p^{LO} \biggl[ 1 \, + \, \delta 
-  \, \frac{1}{400} \, \biggl(\frac{g_A}{\pi F}\biggr)^2 
\int\limits_0^\infty dp \, p^4 \,\, F_{\pi NN}^2(p^2) 
\biggl\{ \frac{21}{w_\pi^3(p^2)} + 
\frac{21}{w_K^3(p^2)} + \frac{4}{w_\eta^3(p^2)} \biggr\} \biggr]\nonumber\\
&\, - \,& \frac{m_N}{50} \, \biggl(\frac{g_A}{\pi F}\biggr)^2 
\int\limits_0^\infty dp \, p^4 \,\, F_{\pi NN}^2(p^2) 
\biggl\{ \frac{11}{w_\pi^4(p^2)} + \frac{1}{w_K^4(p^2)} \biggr\}\nonumber 
\end{eqnarray}
where 
\begin{eqnarray}
\mu_p^{LO} = G_M^p(0)\bigg|_{3q}^{LO} =  
\frac{2m_N \rho R}{1+\frac{3}{2}\rho^2} \nonumber 
\end{eqnarray}  
is the leading-order contribution to the proton magnetic moment. The factor 
\begin{eqnarray}
\delta=-\hat{m}^r \, R \, \rho \, 
\frac{2-\frac{3}{2}\rho^2}{(1+\frac{3}{2}\rho^2)^2} 
\end{eqnarray}  
defines the NLO correction to the nucleon magnetic moments due to the 
modification of the quark w.f. (see Eq. (\ref{GEP_3q})). Note that the 
well-known $SU_6$ relation between nucleon magnetic moments 
$\mu_n/\mu_p = - 2/3$ can be easily deduced from Eq. (\ref{mu_pmu_n}) if: 
i) we restrict to contributions from one-body diagrams in Figs.2a-d,  
corresponding to the additive quark picture, and ii) apply the $SU(3)$-flavor 
limit $(M_\pi = M_K = M_\eta = M_\Phi)$. In particular, we have  
\begin{eqnarray}
\mu_p^{(SU_6)} &\equiv&  - \frac{3}{2} \mu_n^{(SU_6)} = 
\mu_p^{LO} \biggl[ 1 \, + \, \delta -  \, \frac{23}{200} \, 
\biggl(\frac{g_A}{\pi F}\biggr)^2 \int\limits_0^\infty dp \, p^4 \,\, 
\frac{F_{\pi NN}^2(p^2)}{w_\Phi^3(p^2)}\biggr]\nonumber\\ 
&+&\frac{9}{50}\, m_N\,\biggl(\frac{g_A}{\pi F}\biggr)^2\int\limits_0^\infty 
dp \, p^4 \, \frac{F_{\pi NN}^2(p^2)}{w_\Phi^4(p^2)} . 
\end{eqnarray}  
Taking into account the meson-in-flight diagram (Fig.3) generated by two-body 
forces leads to a deviation of the ratio $\mu_n/\mu_p$ from the naive 
$SU_6$ result. 

In general case, for our set of parameters we obtain: 
\begin{eqnarray} 
\mu_p = 2.62 \pm 0.02, \hspace*{.5cm}
\mu_n = -2.02 \pm 0.02, \hspace*{.5cm} \mbox{and} \hspace*{.5cm} 
\frac{\mu_n}{\mu_p} = - 0.76 \pm 0.01 
\end{eqnarray}
where the range of theoretical predictions reflects the variation of the size 
parameter $R$. The separate LO ($\mu_p^{LO}$, $\mu_n^{LO}$) and NLO 
($\mu_p^{NLO}$, $\mu_n^{NLO}$) contributions to the magnetic moments are 
given by 
\begin{eqnarray} 
& &\mu_p^{LO} = 1.8 \pm 0.15 , \hspace*{.5cm}
\mu_n^{LO} \equiv - \frac{2}{3} \mu_p^{LO} ,\\
& &\mu_p^{NLO} = \mu_p - \mu_p^{LO} = 0.82 \pm 0.13, \hspace*{.5cm}
\mu_n^{NLO} = \mu_n - \mu_n^{LO} = - 0.82 \pm 0.08 . \nonumber 
\end{eqnarray}
Our results for $\mu_p$ and $\mu_n$ are close to the ones of the chiral 
soliton quark model by Diakonov and Petrov \cite{Diakonov-Petrov}: 
$\mu_p=2.98$ and $\mu_n=-2.26$. For the electromagnetic nucleon radii 
we obtain  
\begin{eqnarray} 
& &r^p_E = 0.84 \pm 0.05 \, \mbox{fm}, \hspace*{.5cm}
<r^2>^n_E = - 0.046 \pm 0.006 \, \mbox{fm}^2, \\
& &r^p_M = 0.82 \pm 0.02 \, \mbox{fm}, 
\hspace*{.5cm} r^n_M = 0.85 \pm 0.01 \, \mbox{fm} .  \nonumber 
\end{eqnarray}
The LO contributions to the charge radius of the proton 
(see Eq. (\ref{r2ep_LO})) and to the magnetic radii of proton and 
neutron are dominant  
\begin{eqnarray} 
r^p_E\bigg|_{LO} \, = \, 0.77 \pm 0.06 \, \mbox{fm}, \hspace*{.5cm}
r^p_M\bigg|_{LO} \equiv  r^n_M\bigg|_{LO} \, = \, 0.73 \pm 0.06 \, \mbox{fm} . 
\end{eqnarray}
For the neutron charge radius squared we get the observed (negative) sign, 
but its magnitude is smaller than the experimental value. As in the naive 
$SU(6)$ quark model, the LO contribution to the neutron charge radius is zero 
and only one-loop diagrams give nontrivial contributions to this quantity: the 
meson-cloud (MC) $<r^2>^n_{E;{MC}} = -0.072\pm 0.006$ fm$^2$ and 
vertex-correction (VC) $<r^2>^n_{E;{VC}} \approx 0.026$ fm$^2$ diagrams. The 
total contribution to the neutron charge radius is given by 
\begin{eqnarray}
<r^2>^n_E = <r^2>^n_{E;{MC}} + <r^2>^n_{E;{VC}} = -0.046 \pm 0.006 \,\, 
{\rm fm}^2 . 
\end{eqnarray}
The magnitude of our result for the neutron charge radius, which is  
too small compared to the experimental value, is due to the 
reduced contribution of the pion cloud diagram (Fig.2c).  
The size of the meson cloud effects is in turn related to the magnitude 
of the renormalization constant, which for our model is Z=0.9 and 
therefore close to unity. 
With an increased contribution of the pion cloud, that is a decrease of Z,
one is able to reproduce the experimental result for the neutron charge 
radius, but one also obtains a worse description of the magnetic moments 
and the proton charge radius.
Also, perturbation theory up to one loop becomes less reliable in terms
of good convergence properties.
For example, choosing $\rho = 0.12$ (or $g_A=1.5$) and $R=0.52$ fm we obtain
$<r^2>^n_E$ = -0.08 fm$^2$ and $\hat Z=0.73$. However, for the other 
quantities we get a worse fit with: $\mu_p=2.56$, $\mu_n=-2.18$, 
$r^p_E=0.75$ fm, $r^p_M=0.9$ fm and $r^n_M=1$ fm. 
In comparison, the cloudy bag model \cite{Theberge-Thomas} gets a
value of $<r^2>^n_E$ = -0.14 fm$^2$ by an increased contribution of the pion
cloud with $Z \approx 0.73$ \cite{Lu-Thomas-Williams}. As a result,
the predicted values for the proton magnetic moment are also less 
satisfactory with $\mu_p=2.2$ as in the original version of the cloudy bag 
model \cite{Theberge-Thomas} and $\mu_p=2.46$ in the improved 
version \cite{Lu-Thomas-Williams}. The prediction for the proton charge 
radius is also relatively small: $r^p_E$ = 0.73 fm \cite{Theberge-Thomas}. 
 
An improved fit to the neutron charge radius, while keeping the good results
for the other static observables, can possibly be achieved by including
center-of-momentum corrections and also Lorentz boost effects 
Refs. \cite{Theberge-Thomas,Oset,Licht-Pagnamenta,Stanley-Robson,Picek-Tadic,Hwang,Guchon,Kuroda,Fiebig-Hadjimichael,Lubeck-Henley-Wilets}), 
which are based on approximate techniques and are not included yet 
in the present evaluation.
Furthermore, additional contributions of excited quark/antiquark states
in the  meson loop diagrams will further enhance the meson-cloud effects,
which up to now were not studied yet consistently.

In Table I we summarize our results for the static electromagnetic properties 
of the nucleon in comparison with experimental data \cite{PDG}.   
Finally, in Figs.4-8 we indicate the $Q^2$-dependence of the electromagnetic 
form factors.  In Fig.4 we present the result for the proton charge form 
factor at a typical value for the free size parameter $R=0.6 $ fm. The dotted 
line corresponds to the LO result, the short--dashed line marks the NLO 
contribution, while the solid line corresponds to the total result at one 
loop. And, finally, the long-dashed line indicates the dipole fit to 
experimental data given by 
\begin{eqnarray}
G_D(Q^2) = \frac{1}{(1 + Q^2/0.71 \, {\rm GeV}^2)^2} . 
\end{eqnarray}
The dependence of the proton charge form factor on the free parameter $R$ is 
presented in Fig.5. We plot three curves corresponding to $R=0.55$ fm, 
$R=0.6$ fm and $R=0.65$ fm. An increase of $R$ leads to a decrease of the 
proton charge form factor and vice versa. In Fig.6 we give the results 
for the proton charge and the nucleon magnetic form factors for $R=0.6$ fm 
and compare them with the common dipole fit. In Fig.7 we plot our predictions 
for the ratio $\mu_p G_E^p(Q^2)/G_M^p(Q^2)$ at $R=0.6$ fm and compare them 
with experimental data taken from \cite{Jones} and \cite{Milbrath}. 
Our predictions for the ratio $G_M^n(Q^2)/(\mu_n G_D(Q^2))$ are presented in 
Fig.8. Experimental data are taken from \cite{Markovitz}-\cite{Anklin2}. 
Finally, in Fig.9 we plot the results for the neutron charge form factor at 
different values of $R$: $R=0.55$ fm, $R=0.6$ fm and $R=0.65$ fm. 
Our results are lower than the experimental points taken from 
Refs. \cite{Ostrick}-\cite{Platchkov}. 

\section{Summary and conclusions} 

In this paper, we have considered the perturbative chiral quark model (PCQM) 
based on an effective chiral Lagrangian which includes confinement 
phenomenologically. The Lagrangian basically describes nucleons as bound 
states of three valence quarks surrounded by a perturbative cloud of 
pseudoscalar mesons as dictated by the chiral symmetry requirement. 

The main aim of this investigation was to present a consistent formalism 
when treating the Lagrangian in the one-loop expansion. We thereby employed 
a technique, which by appropriate introduction of counterterms allows 
consistent mass renormalization both on the quark and the nucleon level. 
In a further extention of the PCQM we considered the coupling of the photon  
field to the nucleon. Again, charge renormalization was introduced to keep 
a proper definition of the nucleon charge. Local gauge invariance of the 
electromagnetic interaction is fulfilled on the Lagrangian level by 
construction. Due to the noncovariant nature of the effective confinement we 
introduced, local gauge invariance is not necessarily fulfilled for physical 
amplitudes in any reference frame. Only when working in the Breit frame, 
nucleon matrix elements were shown to be consistent with the Ward identity 
without further need to introduce extra terms, like contact interactions, 
in the Lagrangian. 

As one basic application of the developed formalism we considered the 
electromagnetic form factors of the nucleon. Two approximations were 
introduced: 

i) restriction of the relativistic quark states to the ground state, that is 
$N$ and $\Delta$ intermediate states occur in the one-loop terms; 

ii) Gaussian form of the ground state single quark wave function, modelled 
for low-$Q^2$ physics of the calculated observables. 

The derived quantities (magnetic moments, radii and form factors) contain 
only one model parameter $R$, which is related to the radius of the 
three-quark core,  and are otherwise expressed in terms of fundamental 
parameters of low-energy hadron physics: axial coupling constant $g_A$, 
weak pion decay constant $F$ and set of QCD parameters (current quark masses 
$\hat m$ and $m_s$ and quark condensate parameter $B$). Predictions are given 
for a variation of the free parameter $R$ in a physical region from 
0.55 fm to 0.65 fm corresponding to $<r^2_E>^P_{3q-core}$ ranging from 
0.5 fm$^2$ to 0.7 fm$^2$. Our low $Q^2$ results, that is for the nucleon 
magnetic moments, the charge radius of the proton and the magnetic radii of 
the nucleons are in reasonable agreement with data, strengthening the 
phenomenological validity of the PCQM. For the charge radius of 
the neutron we obtain the correct sign but its magnitude is smaller than the 
experimental value. 

We also presented a detailed discussion of the $Q^2$-dependence of the nucleon 
form factors in the space-like region. Although the model underestimates 
the finite $Q^2$ behaviour of the form factors, certain additional physics 
aspects, such as c.m. correction and Lorentz boost effects, are known to 
improve the phenomenological fit (see detailed discussion in 
Refs. \cite{Theberge-Thomas,Oset,Licht-Pagnamenta,Stanley-Robson,Picek-Tadic,Hwang,Guchon,Kuroda,Fiebig-Hadjimichael,Lubeck-Henley-Wilets}). 
The issue of c.m. corrections as evaluated in the PCQM machinery and their 
consistency with the chiral expansion of the nucleon mass will be 
addressed in a forthcoming publication. With the formal and phenomenological 
foundation  established, the proposed model serves as a basis for further 
applications  to recent issues of nucleon structure, for example such as 
photo- and electroproduction of pseudoscalar mesons, which will be pursued 
in future. 

\newpage 
 
{\it Acknowledgements}. This work was supported by the Deutsche 
Forschungsgemeinschaft (DFG, grant FA67/25-1).

\newpage 

%%%%%%%%%%%%%%%%%%%%%%%%%%%%%%%%%%%%%%%%%%%%%%%%%
%            TABLE
%%%%%%%%%%%%%%%%%%%%%%%%%%%%%%%%%%%%%%%%%%%%%%%%%%

\centerline{TABLES}

\vspace*{1cm}

\begin{center}
{\bf Table I.} Static nucleon properties. 

\vspace*{.2cm}

\def\arraystretch{1.5}
\begin{tabular}{|c|c|c|}
\hline
Quantity & Our Approach & Experiment \cite{PDG} \\
\hline
$\mu_{p}$ & 2.62 $\pm$ 0.02 & 2.793 \\
\hline
$\mu_{n}$ & -2.02 $\pm$ 0.02 & -1.913 \\
\hline
$\frac{\displaystyle{\mu_{n}}}{\displaystyle{\mu_{p}}} $ & -0.76 $\pm$ 0.01 
& -0.68 \\
\hline
$r_E^p$ (fm) & 0.84 $\pm$ 0.05 & 0.86 $\pm$ 0.01 \\
\hline
$<r^2>^n_E$ (fm$^2$) & -0.046 $\pm$ 0.006 & -0.119 $\pm$ 0.004 \\
\hline
$r_M^p$ (fm) & 0.82 $\pm$ 0.02 & 0.86 $\pm$ 0.06 \\
\hline
$r_M^n$ (fm) & 0.85 $\pm$ 0.01 & 0.88 $\pm$ 0.07 \\
\hline
\end{tabular}
\end{center}

\newpage

%%%%%%%%%%%%%%%%%%%%%%%%%%%%%%%%%%%%%%%%%%%%%%%%%
%            FIGURES
%%%%%%%%%%%%%%%%%%%%%%%%%%%%%%%%%%%%%%%%%%%%%%%%%%

\begin{figure}[t]
\noindent Fig.1: Diagrams contributing to the baryon energy shift:  

\noindent meson cloud (1a) and exchange diagram (1b).

\vspace*{1cm}
\noindent Fig.2: Diagrams contributing to the nucleon charge: 

\noindent triangle diagram (2a), triangle counterterm diagram (2b), 
meson-cloud diagram (2c), vertex correction diagram (2d), self-energy 
diagrams (2e) and (2f), self-energy counterterm diagrams (2g) and (2h), 
exchange current diagrams (2i) and (2j) and exchange current counterterm 
diagrams (2k) and (2l).

\vspace*{1cm}
\noindent Fig.3: Meson-in-flight diagram.

\vspace*{1cm}
\noindent Fig.4: Proton charge form factor: LO, NLO and Total contributions.

\vspace*{1cm}
\noindent Fig.5: Proton charge form factor for different values of 
the 3q-core radius $R=0.55, 0.6$ and $0.65$ fm.

\vspace*{1cm}
\noindent Fig.6: Proton charge and nucleon magnetic form factors at $R=0.6$ fm.

\vspace*{1cm}
\noindent Fig.7: The ratio $\mu_p G_E^p(Q^2)/G_M^p(Q^2)$ at $R=0.6$ fm. 
Experimental data are taken from \cite{Jones} (JLAB) and 
\cite{Milbrath} (MIT-Bates).

\vspace*{1cm}
\noindent Fig.8: The ratio $G_M^n(Q^2)/(\mu_n G_D(Q^2))$ at $R=0.6$ fm. 
Experimental data are taken from \cite{Markovitz} (MIT-Bates), 
\cite{Anklin1} (NIKHEF), \cite{Bruins} (ELSA) and \cite{Anklin2} (MAMI).

\vspace*{1cm}
\noindent Fig.9: Neutron charge form factor for different values 
of $R=0.55, 0.6$ and $0.65$ fm. 
Experimental data are taken from \cite{Ostrick} (MAMI-1),
\cite{Becker} (MAMI-2), \cite{Meyerhoff} (MAMI-3), \cite{Eden} (MIT) and  
\cite{Platchkov} (PARIS). 
\end{figure}

\newpage 

\begin{figure}
\centering{\
\epsfig{figure=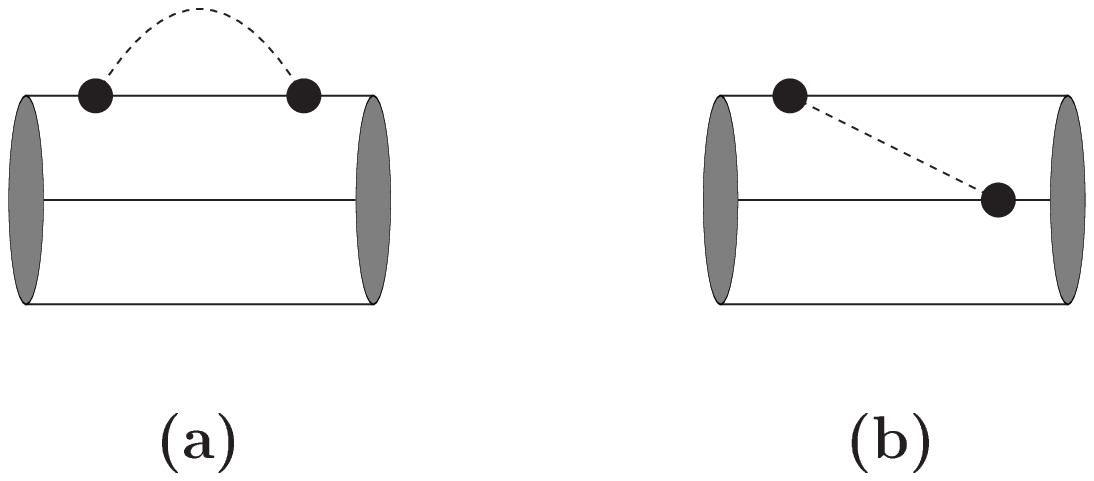,height=21cm}}
\end{figure}

\vspace*{-8cm}

\centerline{\bf Fig.1}

\newpage 

\begin{figure}
\centering{\
\epsfig{figure=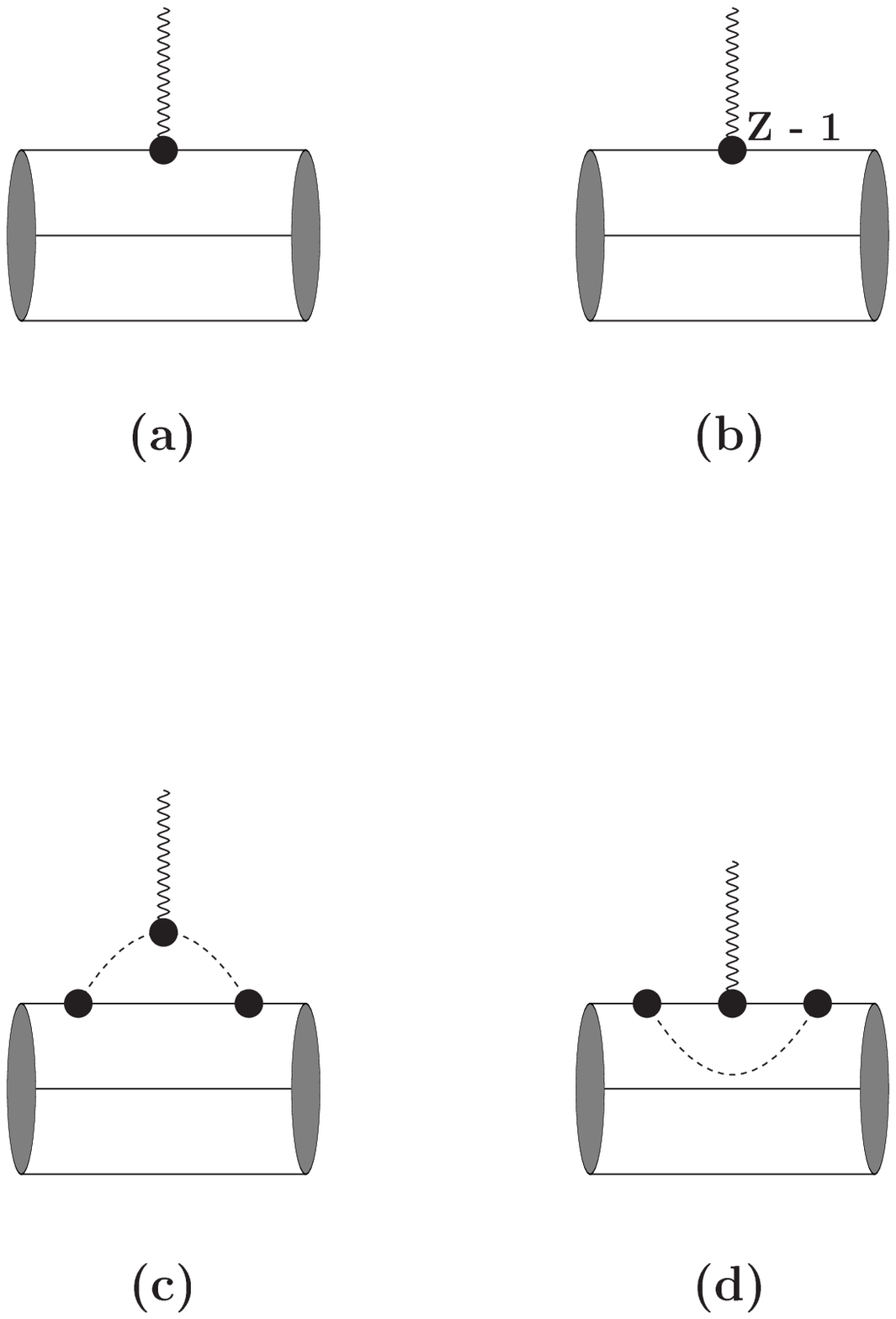,height=21cm}}
\end{figure}

\vspace*{-2cm}

\centerline{\bf Fig.2}

\newpage 

\begin{figure}
\centering{\
\epsfig{figure=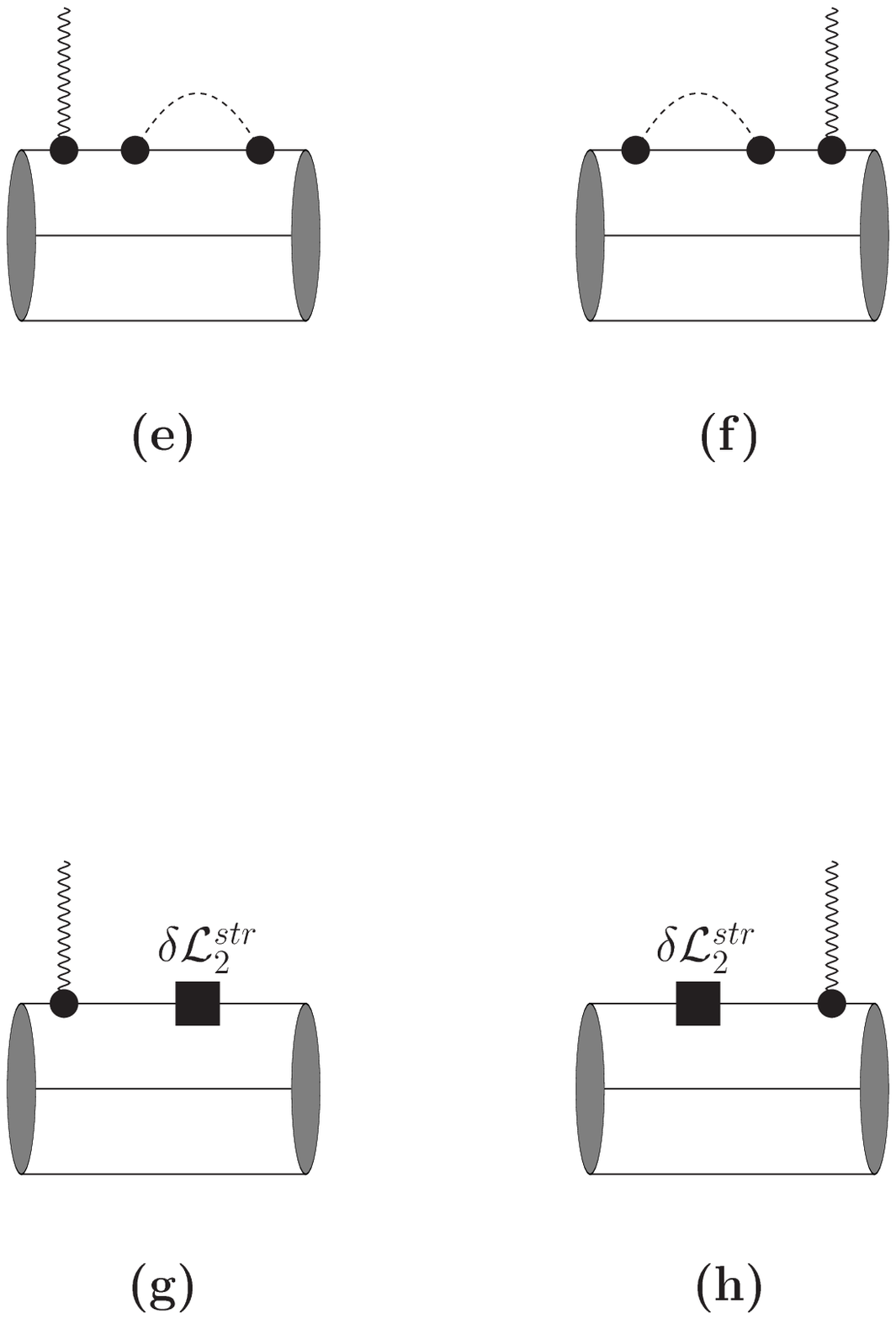,height=21cm}}
\end{figure}

\vspace*{-2cm}

\centerline{\bf Fig.2 (continue)}

\newpage 

\begin{figure}
\centering{\
\epsfig{figure=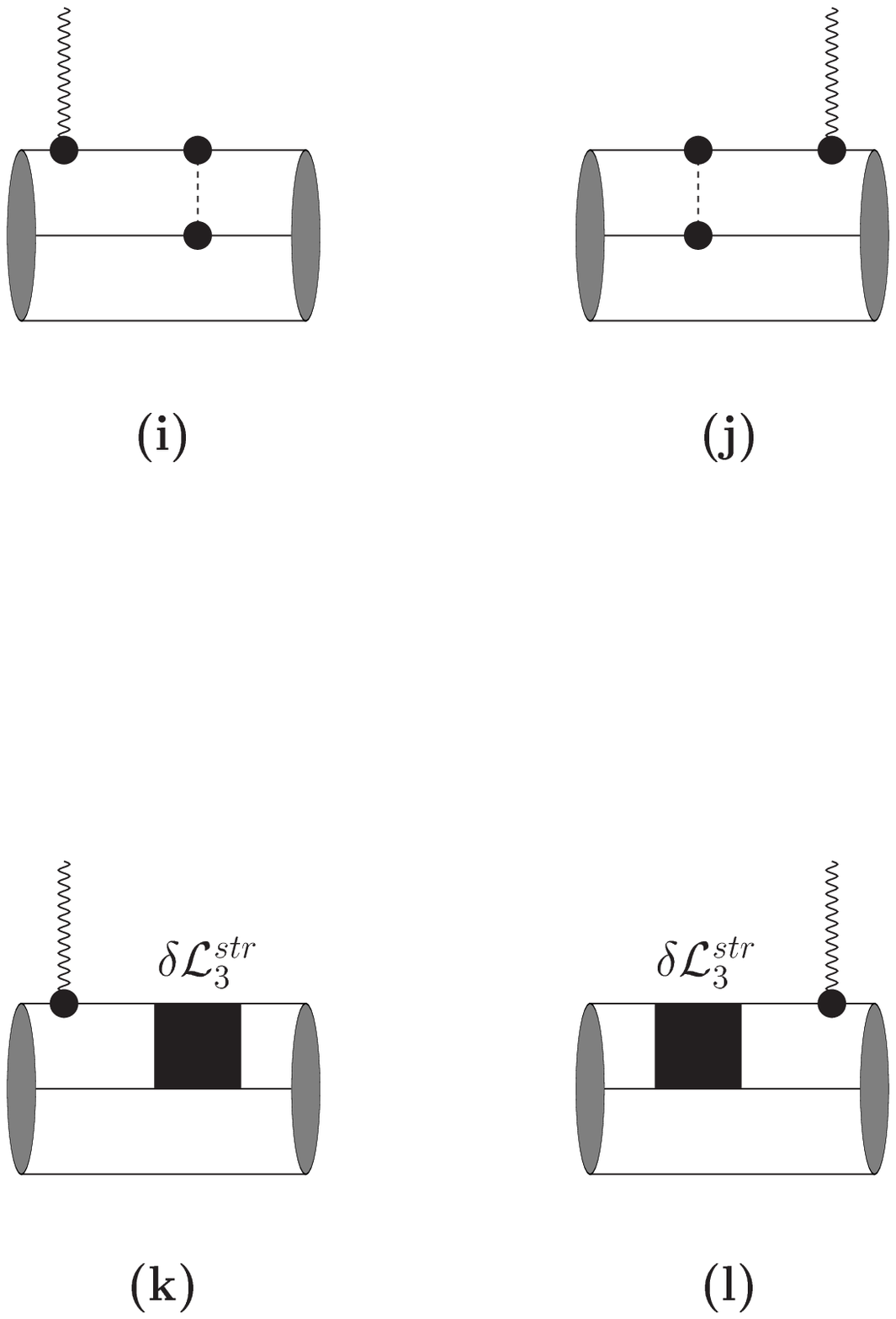,height=21cm}}
\end{figure}

\vspace*{-2cm}

\centerline{\bf Fig.2 (continue)}

\newpage 

\begin{figure}
\centering{\
\epsfig{figure=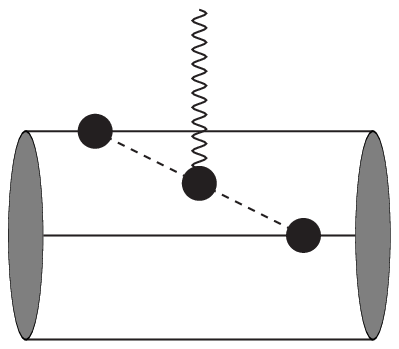,height=21cm}}
\end{figure}

\vspace*{-12cm}

\hspace*{6cm}{\bf Fig.3}

\newpage 

\begin{figure}
\centering{\
\epsfig{figure=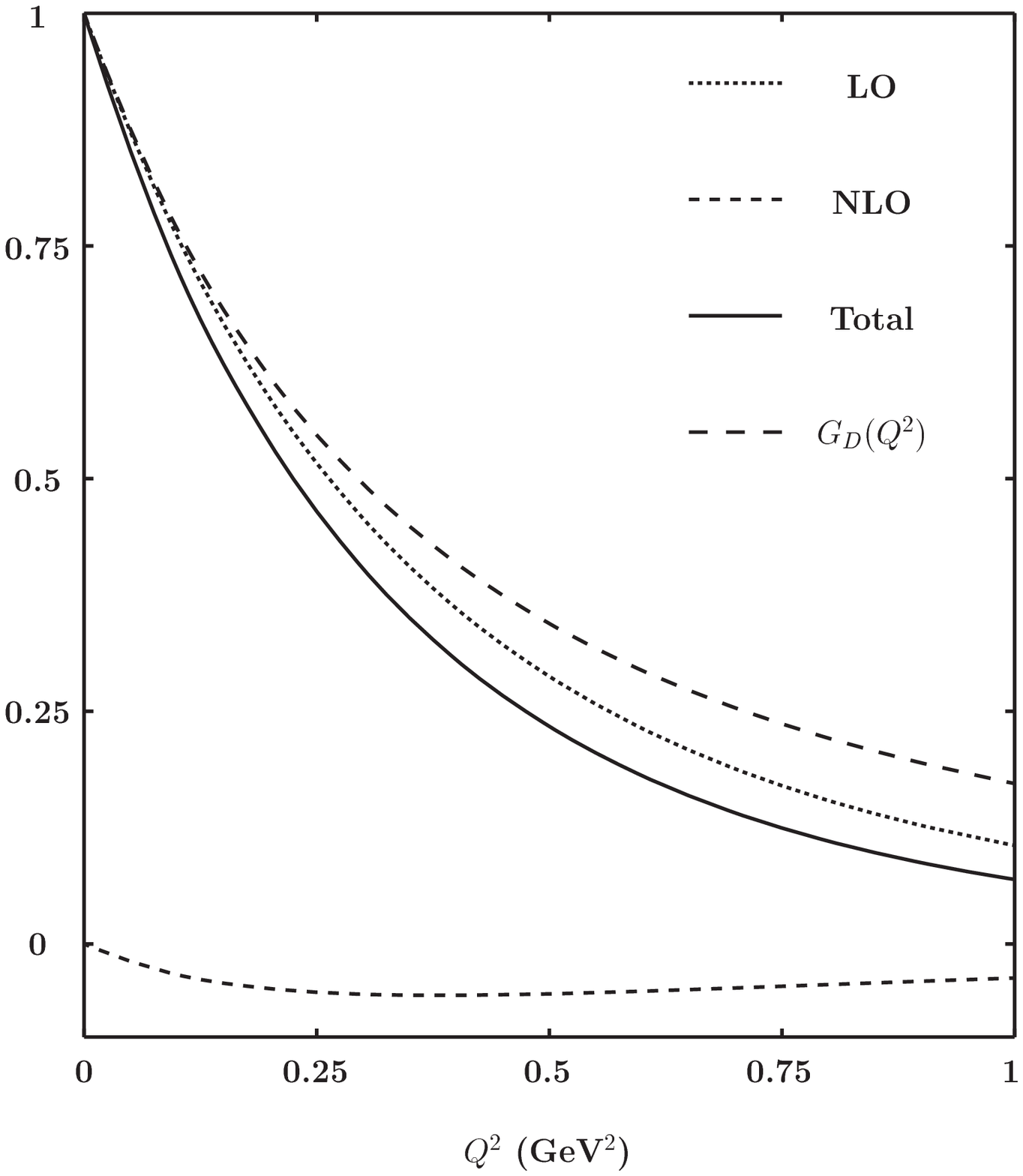,height=21cm}}
\end{figure}

\vspace*{-1cm}

\centerline{\bf Fig.4}

\newpage 

\begin{figure}
\centering{\
\epsfig{figure=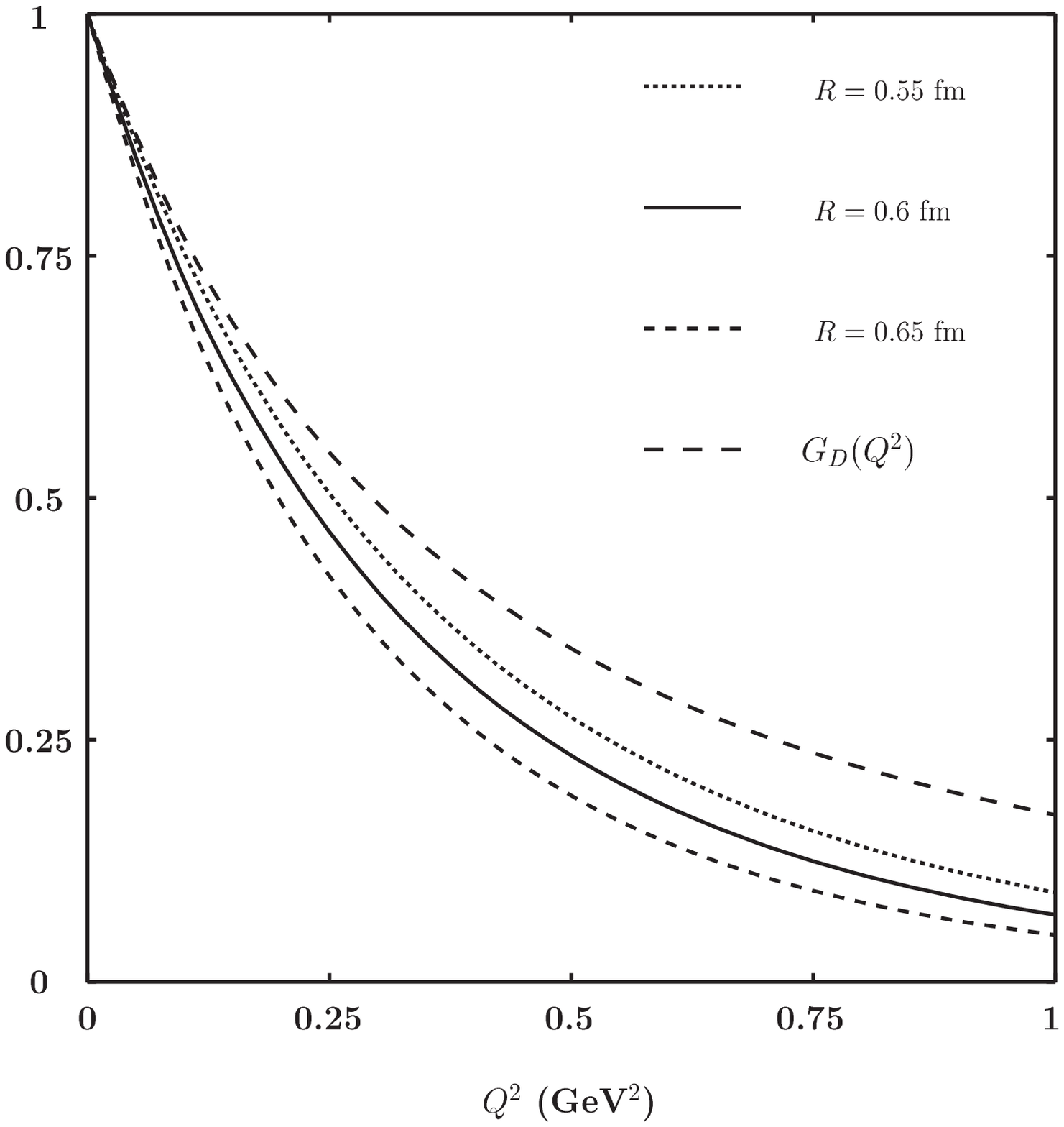,height=21cm}}
\end{figure}

\vspace*{-1cm}

\centerline{\bf Fig.5}

\newpage 

\begin{figure}
\centering{\
\epsfig{figure=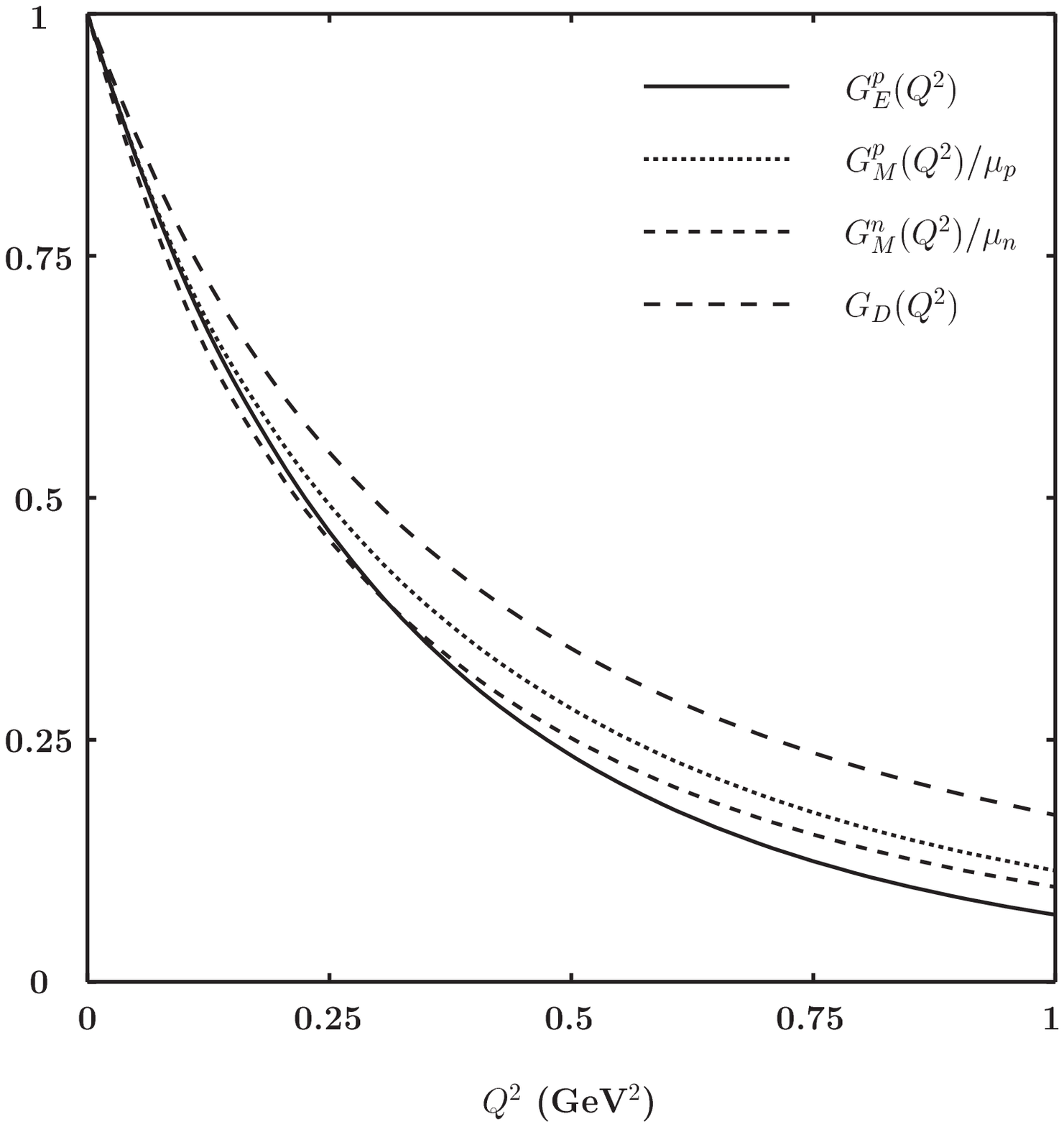,height=21cm}}
\end{figure}

\vspace*{-1cm}

\centerline{\bf Fig.6}

\newpage 

\begin{figure}
\centering{\
\epsfig{figure=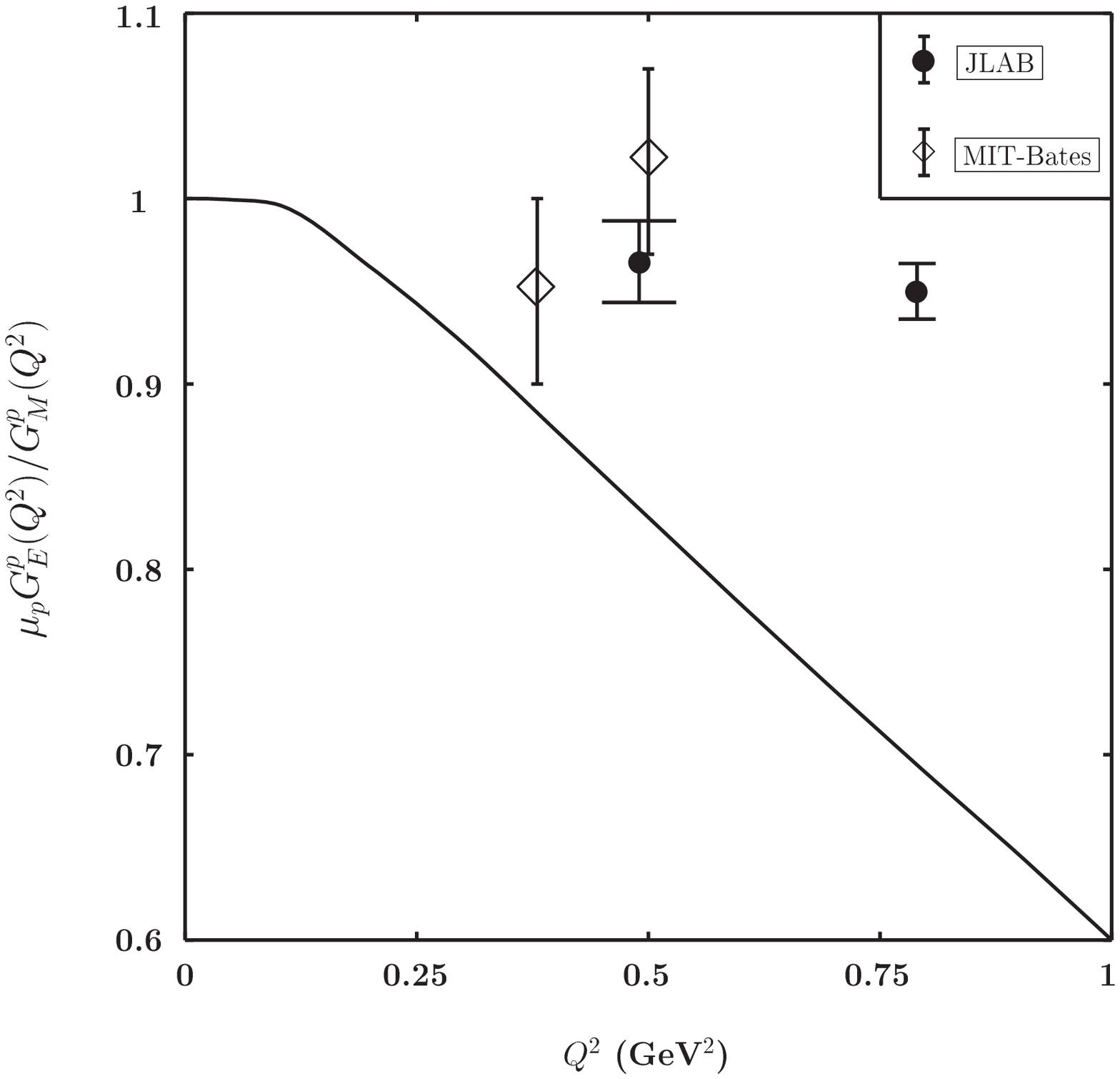,height=21cm}}
\end{figure}

\vspace*{-1cm}

\centerline{\bf Fig.7}

\newpage 

\begin{figure}
\centering{\
\epsfig{figure=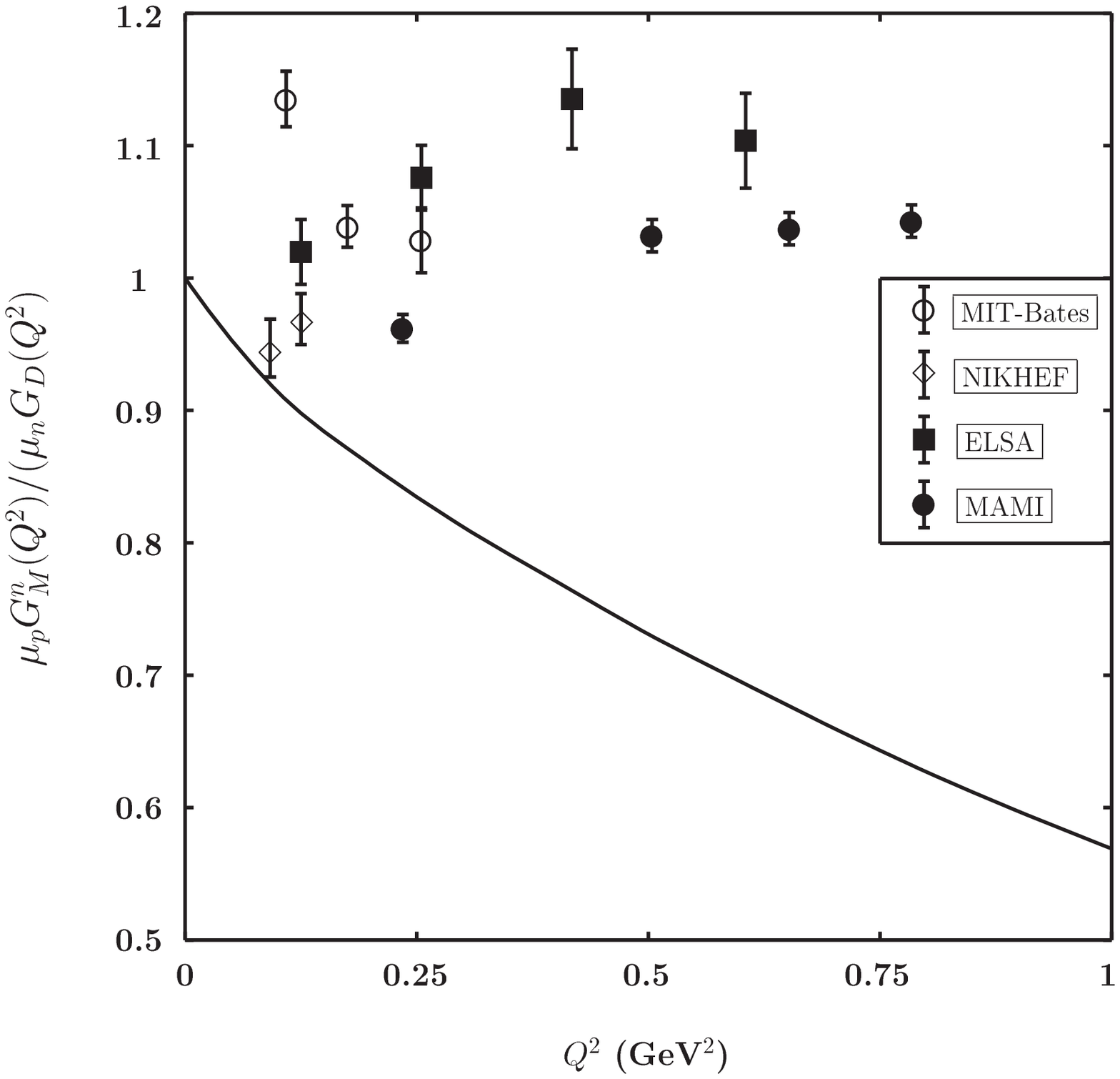,height=21cm}}
\end{figure}

\vspace*{-1cm}

\centerline{\bf Fig.8}

\newpage 

\begin{figure}
\centering{\
\epsfig{figure=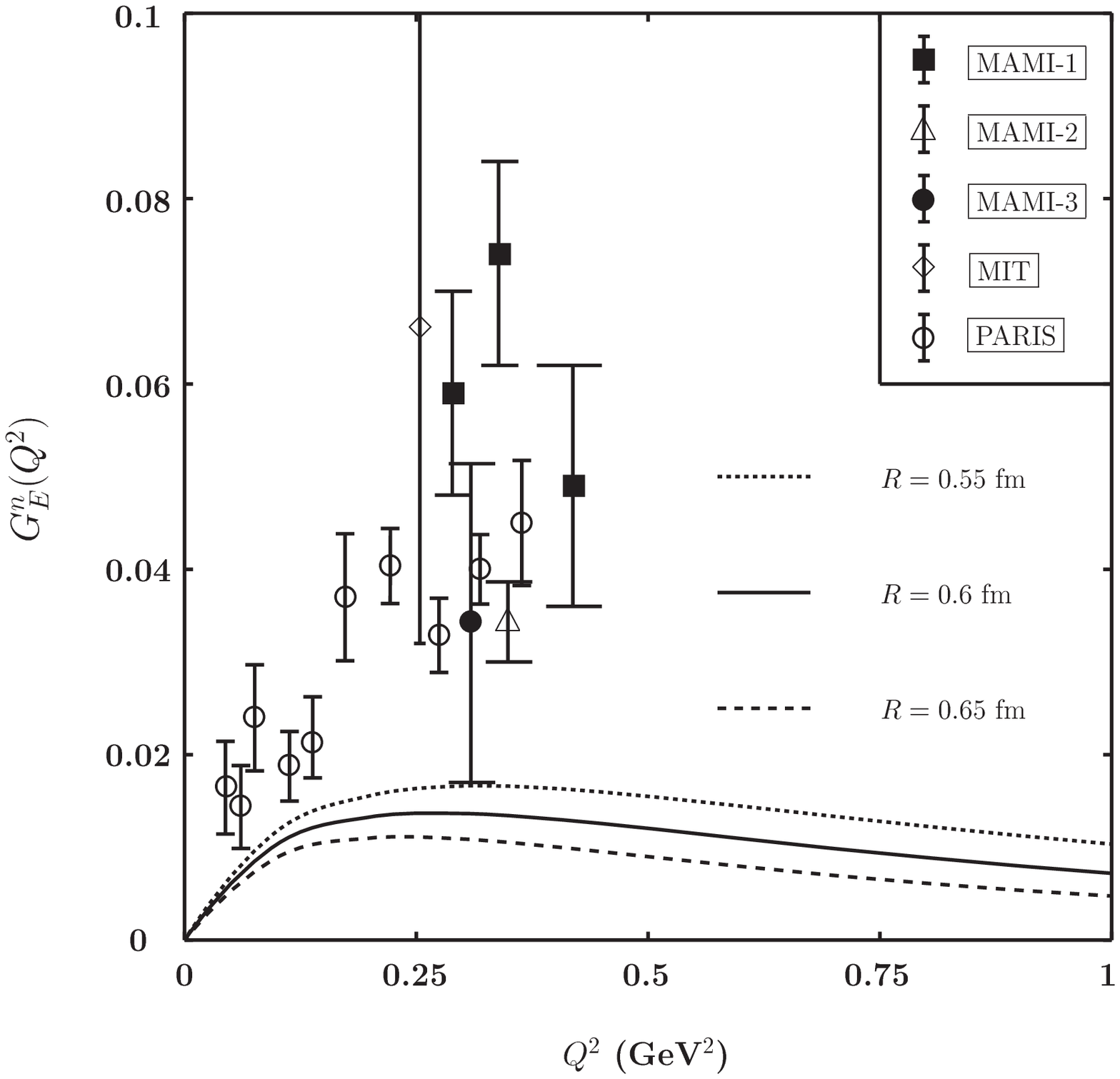,height=21cm}}
\end{figure}

\vspace*{-1cm}

\centerline{\bf Fig.9}

\end{document}